\documentclass[journal,twocolumn]{IEEEtran}
\usepackage{cite}
\usepackage[cmex10]{amsmath}
\usepackage{amsthm}
\usepackage{amssymb}
\usepackage{balance}
\usepackage[final]{graphicx}
\usepackage{color}
\usepackage{epsfig}
\usepackage{epstopdf}
\usepackage{tikz}
\usepackage{hyperref}
\usetikzlibrary{positioning}
\usepackage[ruled,vlined,linesnumbered]{algorithm2e}

\newtheorem{theorem}{Theorem}

\newtheorem{corollary}{Corollary}

\newcommand{\bs}[1]{\boldsymbol{#1}}
\newcommand{\mc}[1]{\mathcal{#1}}

\newcommand{\mb}[1]{\mathbf{#1}}
\newcommand{\mr}[1]{\mathrm{#1}}
\newcommand{\ms}[1]{\mathsf{#1}}
\newcommand{\tr}{\mathrm{Tr}}
\newcommand{\E}{\mathbb{E}}
\newcommand{\lr}[1]{\langle #1 \rangle}
\newcommand{\blr}[1]{\big\langle #1 \big\rangle}

\newcommand{\Blr}[1]{\Big\langle #1 \Big\rangle}

\DeclareMathOperator*{\argmin}{arg\;min}
\DeclareMathOperator*{\argmax}{arg\;max}

\definecolor{LayerColor}{RGB}{230, 230, 230}
\definecolor{InOutColor}{RGB}{240, 243, 255}
\definecolor{cellColor}{RGB}{230, 230, 230}

\begin{document}
	
	\title{A Variational Bayesian Perspective on Massive MIMO Detection}
	
	\author{Duy~H.~N.~Nguyen, Italo~Atzeni, Antti~T\"olli, and A.~Lee~Swindlehurst
		\thanks{Duy~H.~N.~Nguyen is with the Department of Electrical and Computer Engineering, San Diego State University (e-mail: duy.nguyen@sdsu.edu). I.~Atzeni and A.~Tölli are with the Centre for Wireless Communications, University of Oulu (e-mail: \{italo.atzeni, antti.tolli\}@oulu.fi). A.~Lee~Swindlehurst is with the Department of Electrical Engineering and Computer Science, University of California, Irvine (e-mail: swindle@uci.edu).}}
	
	\maketitle
	
	\begin{abstract}
		Optimal data detection in massive multiple-input multiple-output (MIMO) systems requires prohibitive computational complexity. A variety of detection algorithms have been proposed in the literature, offering different trade-offs between complexity and detection performance. In this paper, we build upon variational Bayes (VB) inference to design low-complexity multiuser detection algorithms for massive MIMO systems. We first examine the massive MIMO detection problem with perfect channel state information at the receiver (CSIR) and show that a conventional VB method with known noise variance yields poor detection performance. To address this limitation, we devise two new VB algorithms that use the noise variance and covariance matrix postulated by the algorithms themselves. We further develop the VB framework for massive MIMO detection with imperfect CSIR. Simulation results show that the proposed VB methods achieve significantly lower detection errors compared with existing schemes for a wide range of channel models.
	\end{abstract}
	

	\section{Introduction}
	
	Massive multiple-input multiple-output (MIMO) is a key technology of emerging $5$G networks enabling higher spectral efficiency and improved coverage~\cite{LuLu-JSAC-2014,Rusek-SPMag-2013}. A massive MIMO base station (BS) can concurrently serve a large number of users in the same time-frequency resource via space division multiple access and highly directional beamforming. However, the increased spatial dimension leads to large channel matrices and makes optimal multi-user detection a formidable task. 
	
	The subject of massive MIMO detection has attracted significant interest in recent years, with several contributions offering different trade-offs between computational complexity and detection performance. Conventional detection algorithms based on the maximum a-posteriori (MAP) and maximum likelihood (ML) criteria, which jointly recover all the symbols simultaneously, achieve the optimal detection performance. However, their complexity increases exponentially with the number of users. 
	Linear detectors, such as the matched filter (MF), zero forcing filter, and linear minimum mean squared error (LMMSE) filter, consist of a simple linear pre-processing step to decorrelate the received signals, enabling separate symbol detection on a per-user basis. However, linear detection simply treats the inter-user interference as noise and can thus be highly sub-optimal compared with the MAP/ML detectors, especially in systems with comparable numbers of transmit and receive antennas. Interference cancellation is an attractive alternative solution in terms of both complexity and performance. This family of nonlinear detectors has a long history of development with numerous variants due to the associated design flexibility, see \cite{Yang-Hanzo-CST} and reference therein. The method relies on removing the already detected symbols to facilitate the detection of the remaining ones.  Interference cancellation is thus prone to error propagation, although this issue can be mitigated using soft detected symbols, resulting in the iterative \textit{soft} interference cancellation (SIC) method~\cite{Wang-TCOM-1999,Alexander-TCOM-1999,Choi-Cioffi-WCNC-2000}. Iterative SIC, involving multiple iterations of symbol detection and interference cancellation, can approach the performance of MAP/ML with manageable complexity~\cite{Choi-Cioffi-WCNC-2000,Shlezinger-DeepSIC-TWC-2021}.
	
	Approximate message passing (AMP), originally developed as a computationally efficient algorithm for the recovery of sparse signals~\cite{Donoho}, has also been applied in the context of massive MIMO detection~\cite{Jeon-Studer}. In a MIMO system with independent and identically distributed (i.i.d.) Gaussian channels, AMP decouples the MIMO channel into a set of parallel additive white Gaussian noise (AWGN) channels, thus enabling separate symbol detection. In addition, AMP achieves the minimum symbol error rate (SER) in the large-system limit and shows near-optimal performance for finite-dimensional systems. More importantly, the superior SER performance of AMP can be obtained with very low complexity. The convergence of AMP is established through the algorithm's state evolution for i.i.d. Gaussian~\cite{Bayati-TIT-2011} and i.i.d. sub-Gaussian channels~\cite{Bayati-Annal-2014}. However, AMP may diverge when the channel matrix is ill-conditioned or has non-zero mean. This issue was partially dealt with by the recent development of AMP-based algorithms such as orthogonal AMP (OAMP)~\cite{OAMP-2017} and vector AMP (VAMP)~\cite{VAMP}, which can easily be applied to the MIMO detection problem. 
	It is worth mentioning that rigorous proofs of the state evolution in AMP-based algorithms are generally quite technical and rely on specific assumptions on the channel statistics, e.g., i.i.d. sub-Gaussian or unitarily invariant channels.
	
	Recently, the MIMO detection problem has been tackled using variational Bayes (VB) inference~\cite{Thoota-TCOM-2021}. In uplink MIMO systems, transmitted symbols from the users originate as discrete random variables and are linearly mixed into the received signals. Hence, it is intractable to obtain the marginal posterior distribution of each transmitted symbol. VB inference is a powerful statistical inference framework from machine learning that approximates the intractable posterior distribution of latent variables with a known family of simpler distributions through optimization. Among VB methods, the mean-field approximation enables efficient optimization of the variational distribution over a partition of the latent variables while keeping the variational distributions over the other partitions fixed~\cite{Bishop-2006}. In this paper, we present the VB perspective on the massive MIMO detection problem and compare it with the AMP-based (i.e., AMP and OAMP/VAMP) and SIC methods. We show that mean-field VB decouples a MIMO channel into a set of of parallel AWGN channels. While it is common to assume that the noise variance is known at the receiver, a conventional VB detector relying on knowledge of the noise variance (as in~\cite{Thoota-TCOM-2021}) fails to account for the multiuser interference due to the decoupling of the MIMO channel, and thus results in poor detection performance. We present an analysis of this behavior by connecting the conventional VB detector to the SIC method and make the recommendation to use the noise variance or covariance matrix that is postulated by the VB framework itself. We then develop several new VB algorithms for massive MIMO detection based on closed-form and computationally efficient updates. The resulting iterative algorithms have low complexity comparable to that of AMP-based schemes.
	
	The contributions of this work are listed as follows.
	\begin{itemize}
		\item[$\bullet$] We present a comparison between conventional mean-field VB and SIC methods and provide a new perspective to explain the poor detection performance of the former. We propose two new mean-field VB algorithms for massive MIMO detection with perfect channel state information at the receiver (CSIR) based on the MF (MF-VB) and LMMSE filter (LMMSE-VB), in which the noise variance and covariance matrix are treated as random variables and are thus postulated by the estimation in the algorithms.
		\item[$\bullet$] We further develop MF-VB for the case of imperfect CSIR. The proposed method enables joint estimation of the channel matrix, the symbol vector, and the postulated noise variance.
		\item[$\bullet$] We evaluate the performance of the developed VB algorithms by comparing them with the LMMSE detector as well as the AMP-based and SIC schemes in various channel settings. Numerical results with i.i.d. Gaussian channels indicate that the detection performance of our VB algorithms is better for finite-dimensional systems and comparable to that of SER-optimal AMP-based algorithms in the large-system limit. The proposed VB algorithms, particularly LMMSE-VB, exhibit superior detection performance under correlated channels, realistic 3GPP channels, and channel estimation mismatch.
	\end{itemize}
	
	\emph{Notation.} $x_{ij}$ and $[\mb{X}]_{ij}$ equivalently denote the $(i,j)$th element of a matrix $\mb{X}$; $\mb{x}_i$ is the $i$th column of a matrix $\mb{X}$; $\tr\{\mb{X}\}$ and $|\mb{X}|$ stand for the trace and the determinant, respectively, of a square matrix $\mb{X}$; $\mc{CN}(\bs{\mu},\bs{\Sigma})$ represents a complex Gaussian random vector with mean $\bs{\mu}$ and covariance matrix $\mb{\Sigma}$; $\mc{CN}(\mb{x};\bs{\mu},\mb{\Sigma}) = \big(1/\big(\pi^K|\mb{\Sigma}|\big)\big)\mr{exp}\big(-(\mb{x}-\bs{\mu})^H\mb{\Sigma}^{-1}(\mb{x}-\bs{\mu})\big)$ denotes the probability distribution function (PDF) of a length-$K$ random vector $\mb{x}\sim \mc{CN}(\bs{\mu},\mb{\Sigma})$; $\mathbb{E}_{p(x)}[x]$ and $\mr{Var}_{p(x)}[x]$ are the mean and the variance of $x$ with respect to its distribution $p(x)$; $ \langle x\rangle$, $\big\langle |x|^2 \big\rangle$, and $\sigma_{x}^2=\big\langle|x|^2\big\rangle - \big|\lr{x}\big|^2$ denote the mean, the second moment, and the variance of $x$ with respect to a variational distribution $q(x)$; $\sim$ and $\propto$ stand for ``distributed according to'' and ``proportional to'', respectively.
	
	\emph{Outline.} The rest of the paper is organized as follows. Section~\ref{sec:SM} describes the system model. Section~\ref{sec:conventional-detector} revisits the MIMO detection problem with perfect CSIR, and Section~\ref{sec:VB} presents background on VB inference. Sections~\ref{sec:VB_perf} and~\ref{sec:VB_imperf} propose new VB methods for MIMO detection with perfect and imperfect CSIR, respectively. Section~\ref{sec:simu-Sect} provides numerical results assessing the performance of the proposed algorithms. Finally, Section~\ref{sec:concl} summarizes our contributions.

	\section{System Model} \label{sec:SM}
	
	We consider a MIMO system with $K$ inputs and $M$ outputs, in which the received signal vector $\mb{y} \in \mathbb{C}^M$ is given by
	\begin{eqnarray}\label{system-model}
		\mb{y} = \mb{H}\mb{x} + \mb{n}.
	\end{eqnarray}
	Here, $\mb{H} = [\mb{h}_1,\ldots,\mb{h}_K] \in \mathbb{C}^{M\times K}$ denotes the channel, $\mb{x} = [x_{1}, \ldots, x_{K}]\in \mathbb{C}^K$ is the input signal vector, and $\mb{n} \sim\mc{CN}(\mb{0},N_0\mb{I}_M)$ is the AWGN. Furthermore, we define $\beta = K/M$ as the \emph{system ratio}. Without loss of generality, we refer to the specific case of uplink transmission with $K$ single-antenna users and an $M$-antenna BS.
	We assume that the transmitted symbol $x_i$ from user~$i$ is drawn from a complex-valued discrete constellation $\mc{S}$, e.g., quadrature amplitude modulation (QAM) or phase-shift keying (PSK), and is normalized such that $\mathbb{E}[{x}_i] = 0$ and $\mathbb{E}\big[|{x}_i|^2\big] = 1$. The prior distribution of $x_i$ is given by 
	\begin{eqnarray}\label{x-distribution}
		p(x_i) = \sum_{a\in\mc{S}} p_a\delta(x_i-a),
	\end{eqnarray}
	where $p_a$ corresponds to the known prior probability of the constellation point $a \in \mc{S}$ and $\delta(x_i-a)$ indicates the mass point at $a$. 
	
	Unless otherwise stated, we assume that the channel vector $\mb{h}_i\in\mathbb{C}^M$ associated with user~$i$ is Gaussian with $p(\mb{h}_i) = \mc{CN}(\mb{h}_i;\mb{0},\mb{R}_i)$, where $\mb{R}_i =  \mathbb{E}[\mb{h}_i\mb{h}_i^H]$ is the covariance matrix. Observe that $\mb{R}_i$ is generally not a scaled identity matrix and is typically modeled to reflect the spatial correlation and the large-scale fading from user~$i$ to the BS~\cite{Bjornson-TIT-2014}. Finally, we assume that $\mathbb{E}[\mb{h}_i\mb{h}_j^H] = \mb{0}$ if $i\neq j$. The objective of this paper is to obtain an estimate $\mb{\hat{x}}$ of $\mb{x}$ from the observation $\mb{y}$ with minimum mean squared detection error $\mathbb{E}\big[\|\mb{x}-\mb{\hat{x}}\|^2\big]$.

	\section{MIMO Detection with Perfect CSIR} \label{sec:conventional-detector}
	
	This section revisits MIMO detection with perfect CSIR and describes some state-of-the-art methods that will be used to benchmark the proposed VB algorithms in Section~\ref{sec:simu-Sect}.

	\subsection{Conventional MIMO Detection Schemes}
	
	When the distribution of $\mb{x}$ is discrete, an optimal detector that minimizes the detection error can be obtained through the MAP criterion:
	\begin{eqnarray}
		\mb{\hat{x}}_{\mr{MAP}} &=& \argmax_{\mb{x}\in \mc{S}^K}\; p(\mb{y}|\mb{x};\mb{H})p(\mb{x}) \nonumber \\
		&=& \argmax_{\mb{x}\in \mc{S}^K} \; \big[\ln p(\mb{x}) - N_0^{-1}\|\mb{y}-\mb{Hx}\|^2\big]. 
	\end{eqnarray}
	While the MAP detector (or the ML detector for uniform $p(x_i)$) is optimal in terms of SER, its complexity grows exponentially with the number of inputs, making it infeasible for large-scale MIMO detection. Linear detectors with low complexity are practical candidates for massive MIMO systems. The estimated symbol is obtained via a linear combination of the received signal $\mb{y}$, which is then projected onto the nearest symbol in the constellation $\mathcal{S}$. Among linear detectors, the LMMSE detector achieves the best detection performance. This detector first obtains the LMMSE estimate of $\mb{x}$ as
	\begin{eqnarray}\label{LMMSE}
		\mb{\hat{x}}_{\mr{LMMSE}} = (\mb{H}^H\mb{H} + N_0\mb{I}_{K})^{-1}\mb{H}^H\mb{y},
	\end{eqnarray}
	which is then element-wise projected onto $\mc{S}$. Note that the LMMSE detector requires the inverse of a $(K\times K)$-dimensional matrix. 
	
	\subsection{MIMO Detection via Approximate Message Passing}
	
	The AMP method~\cite{Donoho} was proposed as a computationally efficient iterative method to recover a sparse vector $\mb{x}$ from measurements in the form of \eqref{system-model}. Initializing at iteration $t=1$ with $\hat{x}_i^1 = \E_{p(x_i)}[x_i]$, $\mb{r}^1 = \mb{y}$, and $\nu_1^2 = \mr{Var}_{p({x_i})}[x_i]$, the \textit{\textbf{AMP algorithm}} consists of the following steps:
	\begin{align}
		\mb{z}^t &= \mb{\hat{x}}^t + \mb{H}^H\mb{r}^t, &\textrm{(linear estimator)}&\nonumber \\
		\sigma_t^2 &=N_0 + \beta\nu_t^2, \quad&\textrm{(error variance of $\mb{z}^t$)} &\nonumber \\
		\mb{\hat{x}}^{t+1} &= \eta(\mb{z}^t, \sigma_t^2), \quad&\textrm{(nonlinear denoiser)}&\nonumber \\
		\nu_{t+1}^2 &= \sigma_t^2\big\langle \eta'(\mb{z}^t, \sigma_t^2)\big\rangle, \quad&\textrm{(error variance of $\mb{x}^{t+1}$)} &\nonumber \\
		\mb{r}^{t+1} &= \mb{y} - \mb{H}\mb{\hat{x}}^{t+1} +\beta \frac{\nu_{t+1}^2}{\sigma_t^2}\mb{r}^t, \!\!&\textrm{(Onsager-corrected residual)} &\nonumber 
	\end{align}
	which are repeated until convergence or until a certain number of iterations is reached. Here, $\eta(\cdot,\sigma_t^2): \mathbb{C}^K\rightarrow \mathbb{C}^K$ is a nonlinear \emph{denoiser} parameterized by $\sigma_t^2$ and $\blr{\eta'(\mb{z}^t, \sigma_t^2)} = (1/K) \mr{Tr}\big\{\partial \eta(\mb{z}^t,\sigma_t^2)/\partial \mb{z}^t\big\}$ is its \emph{divergence} at $\mb{z}^t$. When $\mb{H}$ is a large i.i.d. sub-Gaussian matrix, the linear estimator applied to the Onsager-corrected residual decouples the system into $K$ parallel AWGN channels $z_i = x_i + \mc{CN}(0,\sigma_t^2)$~\cite{Donoho,VAMP}. 
	Thus, the denoiser is separable and can be applied element-wise to $\mb{z}^t$. It is worth noting that the postulated error variance $\sigma_t^2$ comprises two terms: one that reflects the true noise variance $N_0$ and one that accounts for the error variance in the denoising step. The AMP method was originally developed for real-valued system models. The complex Bayesian AMP (cB-AMP) algorithm presented above was recently analyzed for massive MIMO detection with complex-valued symbols in~\cite{Jeon-Studer}. The so-called large MIMO AMP (LAMA) algorithm developed therein employs a minimum mean squared error (MMSE) denoiser $\eta(z^t_i,\sigma_t^2) = \ms{F}(z^t_i,\sigma_t^2)$, which is defined as the mean of the posterior distribution $p(x_i|z_i^t;\sigma_t^2)$. Given $\ms{G}(z_i^t,\sigma_t^2)$ as the variance of the posterior distribution $p(x_i|z_i^t;\sigma_t^2)$, the divergence $\blr{\eta'(\mb{z}^t, \sigma_t^2)}$ is equal to $\big(1/(K\sigma_t^2)\big)\sum_{i=1}^K\ms{G}(z^t_i,\sigma_t^2)$~\cite{Rangan-2011}. Effectively, $\nu_{t+1}^2 = \sigma_t^2\blr{\eta'(\mb{z}^t, \sigma_t^2)} = (1/K)\sum_{i=1}^K\ms{G}(z^t_i,\sigma_t^2)$ is the empirical error variance of the MMSE denoiser $\ms{F}(z^t_i,\sigma_t^2)$. 
	
	The convergence of AMP is rigorously established through the algorithm's state evolution for i.i.d. Gaussian and i.i.d. sub-Gaussian $\mb{H}$~\cite{Bayati-TIT-2011,Bayati-Annal-2014}. However, AMP diverges in many practical scenarios
	~\cite{VAMP}. In the context of MIMO detection, AMP diverges when the columns of $\mb{H}$ exhibit correlated elements or when their vector norms are significantly uneven due to users with different large-scale fading coefficients. This limitation prompted the development of AMP-based algorithms that converge for a larger class of matrices than i.i.d. sub-Gaussian, including OAMP~\cite{OAMP-2017} and VAMP~\cite{VAMP} for unitarily invariant matrices. 
	
	While there are subtle differences between OAMP and VAMP in terms of implementation, they are essentially equivalent~\cite{Takeuchi-TIT-2021} as they involve iterations between a linear estimator and a nonlinear denoiser.
	Initializing at iteration $t=1$ with $\hat{x}_1^1 = \E_{p(x_i)}[x_i]$ and $\nu_1^2 = \big(1/\tr\{\mb{H}^H\mb{H}\}\big)\big(\|\mb{y}\|^2 - MN_0\big)$, the \textit{\textbf{OAMP/VAMP algorithm}} consists of the following steps:
	%
	\begin{align}
		\mb{\hat{A}}^t &= \nu_t^2 (\nu_t^2\mb{H}^H\mb{H}+ N_0\mb{I}_K)^{-1}\mb{H}^H, \!\!\!\!\!\!\!& \nonumber \\
		\mb{A}^t &=\frac{K}{\tr\{\mb{\hat{A}}^t\mb{H}\}}\mb{\hat{A}}^t, \!\!\!\!\!\!\!\!\!\!\!\!\!&\textrm{(linear filter)}& \nonumber \\
		\mb{z}^t &= \mb{\hat{x}}^t + \mb{A}^t(\mb{y} - \mb{H}\mb{\hat{x}}^{t}),\quad \!\!\!\!\!\!\!\!\!\!\!\!\!&\textrm{(linear estimator)}&\nonumber \\
		\sigma_t^2 &=\frac{N_0\!\left\|\mb{A}^{t}\right\|_F^2 \! + \! \nu_t^2\!\left\|\mb{I}_K \!-\!\mb{A}^{t}\mb{H}\right\|_F^2}{K},\!\!\!\!\!\!\!\!\!\!\!\!\!& \textrm{(error variance of $\mb{z}^t$)}&\nonumber \\
		\mb{\hat{x}}^{t+1} &= \frac{\eta(\mb{z}^t, \sigma_t^2) - \big\langle {\eta}'(\mb{z}^t, \sigma_t^2)\big\rangle \mb{z}^t}{1 - \big\langle \eta'(\mb{z}^t, \sigma_t^2)\big\rangle}, \!\!\!\!\!\!\!\!\!\!\!\!\!&\textrm{(nonlinear denoiser)}&\nonumber \\
		\nu_{t+1}^2 &= \frac{\|\mb{y}-\mb{H}\mb{\hat{x}}^{t+1}\|^2 - MN_0}{\tr\{\mb{H}^H\mb{H}\}}, \!\!\!\!\!\!\!\!\!\!\!\!\!&\textrm{(error variance of $\mb{x}^{t+1}$)}&\nonumber
	\end{align}
	which are repeated until convergence. In OAMP/VAMP, the linear filter $\mb{\hat{A}}^t$ can be the MF $\mb{H}^H$ (as in AMP), the pseudo-inverse filter $\mb{H}^{\dagger}$, or the LMMSE filter (as in the form presented above). The linear estimate $\mb{z}^t$ is passed through a nonlinear denoiser which also removes the divergence $\blr{\eta'(\mb{z}^t, \sigma_t^2)}$ to obtain a \emph{divergence-free} estimate $\mb{\hat{x}}^{t+1}$. Effectively, OAMP/VAMP decouples the linear MIMO channel into $K$ parallel AWGN channels $z_i = x_i + \mc{CN}(0,\sigma_t^2)$. In its optimal form, OAMP/VAMP adopts the LMMSE filter and the MMSE denoiser, i.e., $\eta(\mb{z}^t,\sigma_t^2) = \ms{F}(\mb{z}^t,\sigma_t^2)$ and $\blr{\eta'(\mb{z}^t, \sigma_t^2)}= \big(1/(K\sigma_t^2)\big)\sum_{i=1}^K\ms{G}(z^t_i,\sigma_t^2)$. Since OAMP/VAMP with the LMMSE filter significantly outperforms its counterparts with MF and the pseudo-inverse filter~\cite{OAMP-2017}, we will only present numerical results for this version and hereafter refer to it simply as OAMP/VAMP. Compared with AMP, OAMP/VAMP requires one matrix inversion per iteration. Interestingly, the matrix inversion in the LMMSE filter can be circumvented in the \emph{economy form} of OAMP/VAMP by first performing the singular value decomposition of the channel. More details on the computation of $p(x_i|z_i^t;\sigma_t^2)$, $\ms{F}(z_i^t, \sigma_t^2)$, and $\ms{G}(z_i^t, \sigma_t^2)$ with discrete $p(x_i)$ used in AMP and OAMP/VAMP are given in Appendix~\ref{sec:append-A}, which also provides an expression for the final MAP estimate~$\hat{x}_i$.
	
	
	\section{Background on VB Inference} \label{sec:VB}
	
	This section gives some background on VB inference, which we will exploit to solve the MIMO detection problem. The goal of VB inference is to find a suitable approximation for a computationally intractable posterior distribution $p(\mb{x}|\mb{y})$ given a probabilistic model that specifies the joint distribution $p(\mb{x},\mb{y})$. Here, $\mb{y}$ represents the set of all the observed variables and $\mb{x}$ is the set of the $m$ latent variables and parameters. 
	
	The VB method consists in finding a distribution $q(\mb{x})$ within a family $\mc{Q}$ of distributions with its own set of variational parameters that make $q(\mb{x})$ as close as possible to the posterior distribution of interest $p(\mb{x}|\mb{y})$. Specifically, VB inference amounts to solving the following~optimization~problem:
	\begin{eqnarray}
		q(\mb{x}) = \argmin_{q(\mb{x}) \in \mc{Q}}\; \mr{KL}\big(q(\mb{x}) \|p(\mb{x}|\mb{y}) \big),
	\end{eqnarray}
	where
	\begin{eqnarray}
		\mr{KL}\big(q(\mb{x}) \|p(\mb{x}|\mb{y}) \big) = \E_{q(\mb{x})} \big[\ln q(\mb{x}) \big] - \E_{q(\mb{x})} \big[\ln p(\mb{x}|\mb{y})\big]
	\end{eqnarray}
	denotes the Kullback-Leibler (KL) divergence of $q(\mb{x})$ from $p(\mb{x}|\mb{y})$. Expanding $p(\mb{x}|\mb{y})$ reveals
	\begin{eqnarray}
		\mr{KL}\big(q(\mb{x}) \|p(\mb{x}|\mb{y}) \big) &=& \E_{q(\mb{x})} \big[\ln q(\mb{x}) \big] - \E_{q(\mb{x})} \big[\ln p(\mb{x},\mb{y})\big] \nonumber \\
		&& + \ln p(\mb{y}). 
	\end{eqnarray}
	Since $p(\mb{y})$ does not depend on $q(\mb{x})$, maximizing the evidence lower bound ($\mr{ELBO}$), 
	defined as
	\begin{eqnarray}
		\mr{ELBO}(q) = \E_{q(\mb{x})} \big[\ln p(\mb{x},\mb{y})\big] - \E_{q(\mb{x})} \big[\ln q(\mb{x}) \big],
	\end{eqnarray}
	is equivalent to minimizing the KL divergence. The maximum possible value of $\mr{ELBO}(q)$ occurs when $q(\mb{x}) = p(\mb{x}|\mb{y})$. 
	
	Since attempting to match the true posterior distribution with an arbitrary $q(\mb{x})$ is typically intractable, it is more practical to consider a restricted family of distributions $q(\mb{x})$. Here, the \emph{mean-field variational family} is constructed such that
	\begin{eqnarray}\label{mean-field}
		q(\mb{x}) = \prod_{i=1}^m q_i(x_i),
	\end{eqnarray}
	where the latent variables are taken to be mutually independent and each is governed by a distinct factor in the variational distribution. Among all distributions $q(\mb{x})$ having the form in \eqref{mean-field}, 
	the general expression for the optimal solution of the variational distribution $q_i(x_i)$ that maximizes the ELBO can be obtained as~\cite{Bishop-2006}
	\begin{eqnarray}
		q_i(x_i) \propto \mr{exp}\left\{\big\langle{\ln p (\mb{y}|\mb{x}) + \ln p(\mb{x})\big\rangle}\right\}, 
	\end{eqnarray}
	where $\lr{\cdot}$ denotes the expectation with respect to all latent variables except $x_i$ using the currently fixed variational distribution $q_{-i}(\mb{x}_{-i}) = \prod_{j\neq i} q_{j}(x_{j})$. By iterating the update of $q_i(x_i)$ sequentially over all $j$, the $\mr{ELBO}(q)$ objective function can be monotonically improved. This is the principle behind the \emph{coordinate ascent variational inference (CAVI)} algorithm, which guarantees convergence to at least a local optimum of $\mr{ELBO}(q)$~\cite{Bishop-2006,Wainwright-2008}.
	
	In the following, we present a theorem on the variational posterior mean of multiple random variables that will be applied later in the paper.
	\begin{theorem}\label{theorem-1} Let the random matrix $\mb{{A}} \in \mathbb{C}^{m\times n}$ and the random vector $\mb{{x}} \in \mathbb{C}^{n}$ be independent with respect to a variational distribution $q(\mb{A},\mb{x})=q(\mb{A})q(\mb{x})$. Assuming that $\mb{{A}}$ is column-wise independent, let $\lr{\mb{a}_i}$ and $\mb{\Sigma}_{\mb{a}_i}$ denote the variational mean and covariance matrix, respectively, of the $i$th column of $\mb{A}$. Furthermore, let $\lr{\mb{x}}$ and $\mb{\Sigma}_{\mb{x}}=\mr{diag}(\sigma_{x_1}^2,\ldots,\sigma_{x_n}^2)$ denote the variational mean and covariance matrix, respectively, of $\mb{x}$. Considering an arbitrary vector $\mb{y} \in \mathbb{C}^{m}$ and defining the expectation $\blr{\|\mb{y}-\mb{A}\mb{x}\|^2}$ with respect to $q(\mb{A},\mb{x})$, we have
		\begin{eqnarray}\label{f-AB}
			\blr{\|\mb{y}-\mb{A}\mb{x}\|^2}& =& \big\|\mb{y}-\lr{\mb{A}}\lr{\mb{x}}\big\|^2 + \tr\big\{\lr{\mb{A}}\mb{\Sigma}_{\mb{x}}\lr{\mb{A}^H}\big\} \nonumber \\
			&& + \sum_{i=1}^n \big\langle|x_i|\big\rangle^2 \tr\{\mb{\Sigma}_{\mb{a}_i}\}.
		\end{eqnarray}
	\end{theorem}
	
	\begin{IEEEproof}
		Expanding $\blr{\|\mb{y}-\mb{A}\mb{x}\|^2}$ and taking into account the independence between $\mb{A}$ and $\mb{x}$, we have 
		\begin{eqnarray}\label{expand}
			\blr{\|\mb{y}-\mb{A}\mb{x}\|^2}&=& \|\mb{y}\|^2 - 2\,\Re\big\{\mb{y}^H\lr{\mb{Ax}}\big\} + \lr{\mb{x}^H\mb{A}^H\mb{A}\mb{x}} \nonumber \\ 
			&=& \big\|\mb{y}-\lr{\mb{A}}\lr{\mb{x}}\big\|^2 - \lr{\mb{x}^H}\lr{\mb{A}^H}\lr{\mb{A}}\lr{\mb{x}} \nonumber \\
			&&+\,\tr\big\{\lr{\mb{A}^H\mb{A}}\lr{\mb{x}\mb{x}^H}\big\}.
		\end{eqnarray}
		Note that $\lr{\mb{x}\mb{x}^H} = \lr{\mb{x}}\lr{\mb{x}^H} + \bs{\Sigma}_{\mb{x}}$. In addition, we have
		\begin{eqnarray}
			\hspace{-3mm} \big[\lr{\mb{A}^H\mb{A}}\big]_{ij} &=& \lr{\mb{a}_i^H\mb{a}_j} \nonumber \\
			&=&
			\left\{ \begin{array}{ll}
				\lr{\mb{a}_i^H}\lr{\mb{a}_i} + \tr\{\mb{\Sigma}_{\mb{a}_i}\}, & \quad \textrm{if } i = j \\
				\lr{\mb{a}_i^H}\lr{\mb{a}_j}, & \quad \textrm{otherwise}.\\
			\end{array} \right.
		\end{eqnarray}
		Thus, it follows that $\lr{\mb{A}^H\mb{A}} = \lr{\mb{A}^H}\lr{\mb{A}} + \mb{D}$, where $\mb{D}= \mr{diag}\big(\tr\{\bs{\Sigma}_{\mb{a}_1}\},\ldots, \tr\{\bs{\Sigma}_{\mb{a}_n}\}\big)$ and, as a result, we have
		\begin{eqnarray}
			\tr\big\{\lr{\mb{A}^H\mb{A}}\lr{\mb{x}\mb{x}^H}\big\} &=& \lr{\mb{x}^H}\lr{\mb{A}^H}\lr{\mb{A}}\lr{\mb{x}}\nonumber \\
			&& + \, \tr\big\{\lr{\mb{A}}\mb{\Sigma}_{\mb{x}}\lr{\mb{A}^H}\big\} \nonumber \\
			&& + \, \lr{\mb{x}^H}\mb{D}\lr{\mb{x}} + \tr\{\mb{D}\mb{\Sigma}_{\mb{x}}\}.
		\end{eqnarray}
		The proof is concluded by removing the duplicated terms in \eqref{expand} and exploiting the fact that $\lr{\mb{x}}^H\mb{D}\lr{\mb{x}} + \tr\{\mb{D}\mb{\Sigma}_{\mb{x}}\} = \sum_{i=1}^n \big|\lr{x_i}\big|^2\tr\{\mb{\Sigma}_{\mb{a}_i}\} + \sum_{i=1}^n \sigma_{x_i}^2\tr\{\mb{\Sigma}_{\mb{a}_i}\} = \sum_{i=1}^n \big\langle|x_i|\big\rangle^2 \tr\{\mb{\Sigma}_{\mb{a}_i}\}$.
	\end{IEEEproof}
	
	\begin{corollary}\label{corollary-1}
		If $\mb{A}$ is deterministic, 
		we have
		\begin{eqnarray}\label{f-x}
			&&\blr{\|\mb{y}-\mb{A}\mb{x}\|^2} 
			= \big\|\mb{y}-\mb{A}\lr{\mb{x}}\big\|^2 +\tr\{\mb{A}\mb{\Sigma}_{\mb{x}}\mb{A}^H\}.
		\end{eqnarray}
	\end{corollary}
	
	\begin{IEEEproof}
		This is a direct result of Theorem~\ref{theorem-1} by exploiting the fact that $\tr\{\mb{\Sigma}_{\mb{a}_i}\}=\mb{0},~\forall i$.
	\end{IEEEproof}
	%
	%
	

	\section{VB Inference for MIMO Detection with Perfect CSIR} \label{sec:VB_perf}
	
	In this section, we apply VB inference to the MIMO detection problem with known channel matrix $\mb{H}$. 
	We first review the conventional mean-field VB framework with known noise variance~\cite{Thoota-TCOM-2021}, which will be referred to in the following as the \textit{\textbf{conv-VB algorithm}}.  We then compare conv-VB with the SIC method~\cite{Choi-Cioffi-WCNC-2000} and identify its limitation with respect to the latter. Lastly, we develop the MF-VB and LMMSE-VB algorithms for MIMO detection.
	
	\begin{figure*}
		\setcounter{equation}{17}
		\begin{eqnarray}\label{g-i}
			g_i(x_i) = \mr{exp}\bigg\{ \ln p(x_i) -N_0^{-1} \bigg(\|\mb{h}_i\|^2|x_i|^2 - 2\,\Re\bigg\{\mb{h}_i^H\bigg(\mb{y} - \sum_{j\neq i}^K \mb{h}_j\lr{x_j}\bigg)x_i^*\bigg\} \bigg) \bigg\}.
		\end{eqnarray}
		\hrulefill
		\setcounter{equation}{15}
	\end{figure*}

	\subsection{Conventional VB Inference with Known Noise Variance}
	
	With known noise statistics, the joint distribution $p(\mb{y},\mb{x};\mb{H},N_0)$ can be factored as
	\begin{eqnarray}
		p(\mb{y},\mb{x};\mb{H},N_0) = p(\mb{y}|\mb{x};\mb{H},N_0)p(\mb{x}),
	\end{eqnarray}
	with $p(\mb{y}|\mb{x};\mb{H},N_0) = \mc{CN}(\mb{y};\mb{Hx},N_0\mb{I}_M)$. Given the observation $\mb{y}$, the mean-field variational distribution $q(\mb{x})= \prod_{i=1}^K q_i(x_i) $ is derived to approximate the posterior distribution 
	$p(\mb{x}|\mb{y};\mb{H},N_0)$. 
	Expanding the conditional $p(\mb{y}|\mb{x};\mb{H},N_0)p(\mb{x})$, taking the expectation with respect to all latent variables except $x_i$ using the variational distribution $\prod_{j\neq i}q_j(x_j)$, and retaining only the components that are related to $x_i$, we have
	\begin{eqnarray} \label{g_s_i}
		q_i(x_i)&\propto& \mr{exp}\big\{\big\langle\ln p(\mb{y}|\mb{x};\mb{H},N_0) + \ln p(\mb{x})\big\rangle\big\} \nonumber \\
		&\propto& \mr{exp}\big\{\big\langle\ln p(\mb{x}) -N_0^{-1}\|\mb{y}-\mb{Hx}\|^2\big\rangle\big\} \nonumber \\
		&=& g_i(x_i),
	\end{eqnarray}
	where $g_i(x_i)$ can be expanded as in \eqref{g-i} at the top of the next page.
	The distribution $q_i(x_i)$ can be normalized such that
	\setcounter{equation}{18}
	\begin{eqnarray}\label{q_s_i}
		q_i(a) = \frac{g_i(a)}{\sum_{b\in \mc{S}} g_i(b)}, \quad \forall a\in \mc{S},
	\end{eqnarray}
	and the variational mean $\lr{x_i}$ with respect to $q_i(x_i)$ is given~by
	\begin{eqnarray}\label{mean_s_i}
		\langle x_i \rangle = \sum_{a\in\mc{S}} a \, q_i(a).
	\end{eqnarray}
	Here, $\langle x_i \rangle$ can be interpreted as a \emph{soft} detection of $x_i$. By iterating the update of $q_i(x_i)$ and $\lr{x_i}$ for $i=1,\ldots,K$, we attain the CAVI algorithm for soft symbol detection. This procedure is summarized in Algorithm~\ref{algo-1}, where $\hat{x}_i^t$ is used to replace the variational mean $\lr{x_i}$ at iteration~$t$. The iterative procedure is repeated until convergence. Note that Algorithm~\ref{algo-1} corresponds to the version of the VB method for MIMO detection developed in~\cite{Thoota-TCOM-2021}. However, we observe in our simulations that Algorithm~\ref{algo-1} occasionally yields a numerical error when the argument inside the exponential function in \eqref{g-i} becomes too large. In the following, we present an equivalent form of the variational distribution $q_i(x_i)$ that helps tackle this issue. 
	
	\begin{algorithm}[t]
		\small
		\textbf{Input:} $\mb{y}$, $\mb{H}$, $N_0$, and prior distributions $\big\{p(x_i)\big\}$\;
		\textbf{Output:} $\mb{\hat{x}}$\; 
		Initialize $\mb{\hat{x}}^1 = \mb{0}$\;
		\SetAlgoNoLine
		\For{$t=1,2,\ldots$}{
			\For{$i=1,2,\ldots,K$}{
				Compute $g_i(x_i)$ as in \eqref{g_s_i}\;
				Normalize the distribution $q_i(x_i)$ as in \eqref{q_s_i}\;
				Compute $\hat{x}_i^t$ as in \eqref{mean_s_i} with respect to $q_i(x_i)$\;}}
		MAP estimate: $\hat{x}_i \leftarrow \argmax_{a\in \mc{S}} q_i(a)$.
		\caption{\textbf{\textit{conv-VB algorithm}} with known noise variance~\cite{Thoota-TCOM-2021}}
		\label{algo-1}
	\end{algorithm}
	
	Let the variational mean $\lr{x_j}$ be the current soft estimate of symbol $x_j$ and denote
	\begin{eqnarray} \label{z-i}
		z_i &=& \frac{\mb{h}_i^H}{\|\mb{h}_i\|^2}\bigg(\mb{y} - \sum_{j\neq i}^K \mb{h}_j\lr{x_j}\bigg). 
	\end{eqnarray}
	Noting that $z_i$ is a constant with respect to the variational distribution $q_i(x_i)$, \eqref{g_s_i}--\eqref{g-i} can be rewritten as
	\begin{eqnarray}\label{inexact-distribution}
		q_i(x_i) 
		&\propto& p(x_i)\,\mr{exp} \big\{-N_0^{-1}\|\mb{h}_i\|^2 \big(|x_i|^2- 2\,\Re\{x_i^* z_i\}\big)\big\} \nonumber \\
		&\propto& p(x_i)\,\mr{exp} \big\{- N_0^{-1} \|\mb{h}_i\|^2 |x_i - z_i|^2\big\} \nonumber \\
		&\propto& p(x_i)\,\mc{CN}(z_i;x_i,\zeta_i^2),
	\end{eqnarray}
	where $\zeta_i^2 = N_0/\|\mb{h}_i\|^{2}$. Here, $\mc{CN}(x_i;z_i,\zeta_i^2)$ can be interpreted as the likelihood function $p(z_i|x_i;\zeta_i^2)$. In other words, the mean-field VB approximation decouples the linear MIMO system into $K$ parallel AWGN channels $z_i = x_i + \mc{CN}\big(0,\zeta_i^2\big)$.
	
	\emph{\textbf{Remark 1:} $z_i$ in \eqref{z-i} can be expressed as $z_i = \lr{x_i} + \big(\mb{h}_i^H/\|\mb{h}_i\|^2\big)\big(\mb{y} - \mb{H}\lr{\mb{x}}\big)$. Interestingly, this expression presents a linear estimator of $x_i$, which is similar to the first step in AMP and in OAMP/VAMP with MF, albeit with the subtle change involving the use of the MF $\mb{h}_i^H/\|\mb{h}_i\|^2$. The variational distribution $q_i(x_i)$ and the corresponding variational mean $\lr{x_i}$ and variance $\sigma_{x_i}^2$ can be obtained in the same manner as $p(x_i|z_i;\zeta_i^2)$ and the corresponding posterior mean $\ms{F}(x_i,\zeta_i^2)$ and variance $\ms{G}(x_i,\zeta_i^2)$ presented in Appendix~\ref{sec:append-A}. Note that $\zeta_i^2$ is now used in place of $\sigma_t^2$ in the AMP-based algorithms. Since now the argument inside the exponential function of $q_i(x_i)$ is always negative, the overflow issue with $q_i(x_i)$ is averted.}
	
	
	\vspace{-0.2cm}
	\subsection{Comparison of Conventional VB Inference with SIC}
	The variational distribution $q_i(x_i)$ in the form of \eqref{inexact-distribution} is the \emph{exact} posterior distribution of the following linear model:
	\vspace{-0.1cm}
	\begin{eqnarray} \label{approx-model}
		\mb{y} = \mb{h}_ix_i + \sum_{j\neq i}^K \mb{h}_j\lr{x_j} + \mb{n},
	\end{eqnarray} 
	where $\sum_{j\neq i}^K \mb{h}_j\lr{x_j}$ is the realized inter-user interference and $\mb{n} \sim \mc{CN}(\mb{0},N_0\mb{I}_M)$. Comparing with the system model \eqref{system-model}, the variational distribution $q_i(x_i)$ corresponds to the true posterior distribution $p(x_i|\mb{y})$ if the estimate $\lr{x_j}$ is the same as the true signal $x_j,~\forall j\neq i$. In this case, the mean-field VB estimation of $x_i$ is Bayes optimal. However, the system model \eqref{system-model} can also be written as
	\vspace{-0.15cm}
	\begin{eqnarray}\label{extract-model}
		\mb{y} = \mb{h}_ix_i + \sum_{j\neq i}^K \mb{h}_j\lr{x_j} + \mb{n}_i, 
	\end{eqnarray}
	where $\mb{n}_i = \sum_{j\neq i}^K \mb{h}_j\big(x_j - \lr{x_j}\big) +\mb{n}$ is the residual interference-plus-noise. With respect to the variational distribution $q_{-i}(\mb{x}_{-i})$, $\mb{n}_i $ has covariance matrix $\mb{C}_{i} = \sum_{j\neq i}^K \sigma_{x_j}^2 \mb{h}_j\mb{h}_j^H + N_0\mb{I}_M$. Since $\mb{C}_{i} \succeq N_0\mb{I}_M$, the variational distribution $q_i(x_i)$ in \eqref{inexact-distribution} may not represent a good approximation of the posterior distribution $p(x_i|\mb{y})$, especially when the residual inter-user interference is not negligible. This observation explains the poor performance of conv-VB with respect to the AMP-based algorithms. Note that this issue was previously reported in~\cite{Krzakala-ISIT-2014} for the sparse signal recovery problem by comparing the difference between the fixed points of AMP and the mean-field VB schemes.
	
	We now compare conv-VB with SIC-based MIMO detection~\cite{Choi-Cioffi-WCNC-2000}. With a slight abuse of notation, we use $\lr{x_i}$ to denote the current soft estimate of $x_i$ in SIC. The key idea of SIC is to first approximate the residual interference-plus-noise ${\mb{n}}_i$ as Gaussian while fixing the estimate $\lr{x_j},~\forall j\neq i$. Then, the likelihood function $p\big(\mb{y}|x_i;\lr{\mb{x}_{-i}}\big)$ is approximated as $\mc{CN}\big(\mb{y}; \mb{h}_ix_i + \sum_{j\neq i}^K \mb{h}_j\lr{x_j}, \mb{C}_{i}\big)$, enabling a tractable derivation of the posterior distribution $p\big(x_i|\mb{y};\lr{\mb{x}_{-i}}\big)$ via the Bayes' theorem.\footnote{It can be proved that $\mb{n}_i$ is Gaussian for sufficiently large $K$ using Lindeberg's condition for the central limit theorem
		~\cite{Feller-1968}. In addition, the covariance matrix $\mb{C}_i$ of $\mb{n}_i$ for an i.i.d. channel $\mb{H}$ and sufficiently large $K$ tends to a scaled identity matrix, making $\mb{n}_i$ i.i.d. as well.} To this end, we examine two approaches to combine the output signal $\mb{y}$, cancel the interference $\sum_{j\neq i}^K \mb{h}_j\lr{x_j}$, and generate the soft estimate $\lr{x_i}$.
	\begin{itemize}
		\item[$\bullet$] SIC with MF (\textbf{\textit{MF-SIC algorithm}}): By applying the MF $\mb{h}_i^H/\|\mb{h}_i\|^2$ as in \eqref{z-i} to \eqref{extract-model}, one attains a linear estimate $z_i$ of $x_i$ with interference cancellation as $z_i \approx x_i + \mc{CN}\big(0,\mb{h}_i^H\mb{C}_i\mb{h}_i/\|\mb{h}_i\|^4\big)$. The soft estimate $\lr{x_i}$ can then be obtained as the posterior mean of $p\big(x_i|z_i;\mb{h}_i^H\mb{C}_i\mb{h}_i/\|\mb{h}_i\|^4\big)$.
		\item[$\bullet$] SIC with LMMSE filter (\textbf{\textit{LMMSE-SIC algorithm}}): Observe that 
		\begin{eqnarray}
			&&p\big(\mb{y}|x_i;\lr{\mb{x}_{-i}}\big) \nonumber \\
			&&\approx\mc{CN}\bigg(\mb{y}; \mb{h}_ix_i + \sum_{j\neq i}^K \mb{h}_j\lr{x_j}, \mb{C}_{i}\bigg) \nonumber \\
			&&= \mc{CN}\bigg(x_i;\frac{\mb{h}_i^H\mb{C}_i^{-1}\big(\mb{y}-\sum_{j\neq i}^K \mb{h}_j\lr{x_j} \big)}{\mb{h}_i^H\mb{C}_i^{-1}\mb{h}_i},\frac{1}{\mb{h}_i^H\mb{C}_i^{-1}\mb{h}_i}\bigg) \nonumber \\
			&&=\mc{CN}\big(z_i;x_i, {1}/{(\mb{h}_i^H\mb{C}_i^{-1}\mb{h}_i)}\big),
		\end{eqnarray}
		where
		\begin{eqnarray}\label{SIC-MMSE}
			z_i = \frac{\mb{h}_i^H\mb{C}_i^{-1}\big(\mb{y}-\sum_{j\neq i}^K \mb{h}_j\lr{x_j} \big)}{\mb{h}_i^H\mb{C}_i^{-1}\mb{h}_i}.
		\end{eqnarray}
		Thus, we have $z_i \approx x_i + \mc{CN}\big(0,1/(\mb{h}_i^H\mb{C}_i^{-1}\mb{h}_i)\big)$. The soft estimate $\lr{x_i}$ can then be obtained as the mean of the posterior distribution $p\big(x_i|z_i;1/(\mb{h}_i^H\mb{C}_i^{-1}\mb{h}_i)\big)$. 
		Observe that $z_i$ in the form of \eqref{SIC-MMSE} is the LMMSE estimate of $x_i$ after canceling the inter-user interference and whitening with the residual interference-plus-noise covariance matrix $\mb{C}_i$.
	\end{itemize}
	
	MF-SIC and LMMSE-SIC proceed with the iterative update over $\{x_i\}$ until convergence. Except for the removal of the divergence terms, the implementation of MF-SIC and LMMSE-SIC is similar to that of AMP and OAMP/VAMP, respectively. However, while SIC methods yield a good detection performance (as shown in Section~\ref{sec:simu-Sect}), they are characterized by two major shortcomings. First, the $K$ $(M \times M)$-dimensional residual interference-plus-noise covariance matrices $\{\mb{C}_i\}$ need to be computed (and inverted in LMMSE-SIC) at each iteration, which leads to prohibitive complexity for large systems. Second, the convergence of SIC-based MIMO detection has not been formally established. To address the poor performance of conv-VB with known noise variance and the shortcomings of the SIC algorithms, we develop two novel VB schemes that jointly estimate the symbol vector and the postulated noise variance or covariance matrix.

	\subsection{Proposed MF-VB for MIMO Detection} \label{sec:noise-variance-Sect}
	
	In practice, the noise variance $N_0$ is not known \emph{a priori} and needs to be estimated as well. Moreover, the conventional VB method with known noise variance does not take into account the residual inter-user interference. Here, we consider the residual interference-plus-noise as a random variable $N_0^{\mathrm{post}}$, which is postulated by the estimation in the VB framework. For ease of computation, we use $\gamma = 1/N_0^{\mathrm{post}}$ to denote the precision of the estimation. We assume a conjugate prior Gamma distribution $\mr{Gamma}(a_0,b_0)$ for $\gamma$, where $a_0$ and $b_0$ are the shape and rate parameters of the distribution, respectively. The PDF of $\gamma$ is thus given by
	\begin{eqnarray}
		p(\gamma) = \frac{b_0^{\gamma}}{\Gamma(a_0)} \gamma^{a_0-1}\mr{e}^{-b_0 \gamma},
	\end{eqnarray}
	where $\Gamma(a_0)$ is the Gamma function. 
	Treating the precision $\gamma$ as a random variable, the joint distribution $p(\mb{y},\mb{x},\gamma;\mb{H})$ can be factored as
	\begin{eqnarray}\label{factor-gamma}
		p(\mb{y},\mb{x},\gamma;\mb{H}) = p(\mb{y}|\mb{x},\gamma;\mb{H})p(\mb{x})p(\gamma),
	\end{eqnarray}
	where $p(\mb{y}|\mb{x},\gamma;\mb{H}) = \mc{CN}(\mb{y};\mb{Hx},\gamma^{-1}\mb{I}_M)$. Given the observation $\mb{y}$, we aim at obtaining the mean-field variational distribution $q(\mb{x},\gamma)$ such that
	\vspace{-0.15cm}
	\begin{eqnarray}
		p(\mb{x},\gamma|\mb{y};\mb{H}) \approx q(\mb{x},\gamma) =\prod_{i=1}^K q_i(x_i)q(\gamma).
	\end{eqnarray}
	The optimization of $q(\mb{x},\gamma)$ is executed by iteratively updating $\{x_i\}$ and $\gamma$ as follows.
	
	\textit{1) Updating $x_i$.} The variational distribution $q_i(x_i)$ is obtained by expanding the conditional in \eqref{factor-gamma} and taking the expectation with respect to all latent variables except $x_i$ using the variational distribution $\prod_{j\neq i}^K q_j(x_j)q(\gamma)$:
	\vspace{-0.1cm}
	\begin{eqnarray} \label{q-x-MF-VB}
		q_i(x_i)&\propto& \mr{exp}\big\{\big\langle\ln p(\mb{y}|\mb{x},\gamma;\mb{H}) + \ln p(\mb{x}) \big\rangle\big\} \nonumber \\
		&\propto& \mr{exp}\big\{\blr{\ln p(\mb{x}) -\gamma\|\mb{y}-\mb{Hx}\|^2}\big\} \nonumber\\
		&\propto& p(x_i)\,\mr{exp} \big\{- \lr{\gamma}\|\mb{h}_i\|^2 |x_i - z_i|^2\big\} \nonumber \\
		&\propto & p(x_i)\,\mc{CN}\big(z_i;x_i,{1}/{\big(\lr{\gamma}\|\mb{h}_i\|^2\big)}\big),
	\end{eqnarray}
	where $z_i$ is the linear estimate of $x_i$ as defined in \eqref{z-i}. The variational distribution $q_i(x_i)$ can be easily realized by normalizing $p(x_i)\,\mc{CN}\big(z_i;x_i,1/\big(\lr{\gamma}\|\mb{h}_i\|^2\big)\big)$. The variational mean and variance are then computed as $\lr{x_i} = \ms{F}\big(z_i,1/\big(\lr{\gamma}\|\mb{h}_i\|^2\big)\big)$ and $\sigma_{x_i}^2 = \ms{G}\big(z_i,1/\big(\lr{\gamma}\|\mb{h}_i\|^2\big)\big)$, respectively. 
	
	\begin{figure*}
		\setcounter{equation}{38}
		\begin{eqnarray}
			q(\mb{W}) &\propto& \mr{exp}\big\{\ln|\mb{W}|-\big\langle (\mb{y}-\mb{Hx})^H\mb{W}(\mb{y}-\mb{Hx}) \big\rangle +(n-M)\ln|\mb{W}| - \tr\{\mb{W}_0^{-1}\mb{W}\}\big\} \nonumber \\
			&\propto& \mr{exp}\Big\{(n-M+1)\ln|\mb{W}| -\tr\Big\{\big(\mb{W}_0^{-1} + \big(\mb{y}-\mb{H}\lr{\mb{x}}\big)\big(\mb{y}-\mb{H}\lr{\mb{x}}\big)^H+\mb{H}\mb{\Sigma}_{\mb{x}}\mb{H}^H\big)\mb{W}\Big\}\Big\}. \label{W-distribution} \\
			\lr{\mb{W}} &=& (n+1) \big(\mb{W}_0^{-1} + \big(\mb{y}-\mb{H}\lr{\mb{x}}\big)\big(\mb{y}-\mb{H}\lr{\mb{x}}\big)^H+\mb{H}\mb{\Sigma}_{\mb{x}}\mb{H}^H\big)^{-1}. \label{W-mean}
		\end{eqnarray}
		\hrulefill
		\setcounter{equation}{30}
	\end{figure*}
	
	\textit{2) Updating $\gamma$.} The variational distribution $q(\gamma)$ is obtained by taking the expectation of the conditional in \eqref{factor-gamma} with respect to $q(\mb{x})$:
	\vspace{-0.1cm}
	\begin{eqnarray} 
		q(\gamma) 
		&\propto&\mr{exp}\big\{ \big\langle \ln p(\mb{y}|\mb{x},\gamma;\mb{H}) + \ln p(\gamma)\big\rangle\big\} \nonumber \\
		&\propto& \mr{exp}\big\{ M\ln \gamma \! - \! \gamma \blr{\|\mb{y}- \mb{Hx}\|^2} \! + \! (a_0-1)\ln\gamma \! - \! b_0\gamma \big\} \nonumber \\
		&\propto& \mr{exp}\big\{ (a_0 + M - 1)\ln \gamma \nonumber \\
		&& - \, \gamma\big(b_0+\big\|\mb{y} - \mb{H}\lr{\mb{x}}\big\|^2+ \tr\{\mb{H}\mb{\Sigma}_{\mb{x}}\mb{H}^H\} \big)\big\}.
	\end{eqnarray}
	Note that the last step is obtained as a result of Corollary~\ref{corollary-1}. The variational distribution $q(\gamma)$ is thus Gamma with mean
	\vspace{-0.1cm}
	\begin{eqnarray} \label{gamma-update}
		\lr{\gamma} = \frac{a_0 + M }{b_0 + \big\|\mb{y} - \mb{H}\lr{\mb{x}}\big\|^2+ \tr\{\mb{H}\mb{\Sigma}_{\mb{x}}\mb{H}^H\} }.
	\end{eqnarray}
	
	By iteratively optimizing $\big\{q_i(x_i)\big\}$ and $q(\gamma)$, we obtain the CAVI algorithm for estimating $\mb{x}$ and the precision $\gamma$. We refer to this scheme as the \textbf{\textit{MF-VB algorithm}} due to the use of the MF $\mb{h}_i^H/\|\mb{h}_i\|^2$ to obtain the linear estimate $z_i$ in \eqref{z-i}. 
	
	\emph{\textbf{Remark 2:} If the improper prior $\mr{Gamma}(0,0)$ is used, $1/\lr{\gamma} = (1/M)\big(\big\|\mb{y} - \mb{H}\lr{\mb{x}}\big\|^2+ \tr\{\mb{H}\mb{\Sigma}_{\mb{x}}\mb{H}^H\} \big)$ is now the point estimate of the deterministic unknown $N_0^{\mathrm{post}}$. Similar to the AMP-based algorithms, the term $\big\|\mb{y} - \mb{H}\lr{\mb{x}}\big\|^2$ reflects the empirical estimate of the true noise variance $N_0$, whereas the term $\tr\{\mb{H}\mb{\Sigma}_{\mb{x}}\mb{H}^H\}$ reflects the empirical error variance induced by the MMSE denoiser $\ms{F}\big(z_i,1/\big(\lr{\gamma}\|\mb{h}_i\|^2\big)\big)$. This result, corresponding to the M-step in variational Bayesian expectation-maximization~\cite{Wainwright-2008}, coincides with the estimated noise variance proposed in~\cite{Krzakala-ISIT-2014}. However, unlike~\cite{Krzakala-ISIT-2014}, we treat the reciprocal of the postulated noise variance as a random variable under the VB framework.}
	
	\emph{\textbf{Remark 3:} If $N_0$ is known, a weakly informative Gamma prior, e.g., $\mr{Gamma}(1,N_0)$ that results in $\mathbb{E}[\gamma] = 1/N_0$, can be used instead of $\mr{Gamma}(0,0)$. Interestingly, we observe through numerous simulations that the MF-VB approach based on the improper prior $\mr{Gamma}(0,0)$ yields a marginally more accurate estimation of $\mb{x}$ than the one based on a weakly informative prior. Thus, we use the improper prior $\mr{Gamma}(0,0)$ in the numerical results for MF-VB.}
	
	\vspace{-0.2cm}
	\subsection{Proposed LMMSE-VB for MIMO Detection}
	
	We now develop a VB method to estimate the input signal $\mb{x}$ using the postulated noise covariance matrix $\mb{C}^{\mathrm{post}}$ instead of the postulated noise variance $N_0^{\mathrm{post}}$ as in Section~\ref{sec:noise-variance-Sect}. For ease of computation, we use $\mb{W} = (\mb{C}^{\mathrm{post}})^{-1}$ to denote the precision matrix and assume a conjugate prior complex Wishart distribution $\mc{CW}(\mb{W}_0,n)$ for $\mb{W}$, where $\mb{W}_0\succeq \mb{0}$ is the scale matrix and $n\geq M$ indicates the number of degrees of freedom. The PDF of $\mb{W}\sim \mc{CW}(\mb{W}_0,n)$ satisfies 
	\vspace{-0.1cm}
	\begin{eqnarray}
		p(\mb{W}) \propto |\mb{W}|^{n-M}\mr{exp}\big(-\tr\{\mb{W}_0^{-1}\mb{W}\}\big).
	\end{eqnarray} 
	The joint distribution $p(\mb{y},\mb{x},\mb{W};\mb{H})$ can be factored as
	\vspace{-0.1cm}
	\begin{eqnarray}\label{factor-W}
		p(\mb{y},\mb{x},\mb{W};\mb{H}) = p(\mb{y}|\mb{x},\mb{W};\mb{H})p(\mb{x})p(\mb{W}),
	\end{eqnarray}
	where $p(\mb{y}|\mb{x},\mb{W};\mb{H}) = \mc{CN}(\mb{y};\mb{Hx},\mb{W}^{-1})$. Given the observation $\mb{y}$, we aim at obtaining the mean-field variational distribution $q(\mb{x},\mb{W})$ such that
	\vspace{-0.15cm}
	\begin{eqnarray}
		p(\mb{x},\mb{W}|\mb{y};\mb{H}) \approx q(\mb{x},\mb{W}) = \prod_{i=1}^K q_i(x_i)q(\mb{W}).
	\end{eqnarray}
	The optimization of $q(\mb{x},\mb{W})$ is executed by iteratively updating $\{x_i\}$ and $\mb{W}$ as follows.
	
	\textit{1) Updating $x_i$.} The variational distribution $q_i(x_i)$ is obtained by expanding the conditional in \eqref{factor-W} and taking the expectation with respect to all latent variables except $x_i$ using the variational distribution $\prod_{j\neq i}^K q_j(x_j)q(\mb{W})$:
	\vspace{-0.1cm}
	\begin{eqnarray} \label{q-x-LMMSE-VB}
		q_i(x_i)&\propto& \mr{exp}\big\{\big\langle\ln p(\mb{y}|\mb{x},\mb{W};\mb{H}) + \ln p(\mb{x}) \big\rangle\big\} \nonumber \\
		&\propto& \mr{exp}\big\{\big\langle\ln p(\mb{x}) -(\mb{y}-\mb{Hx})^H\mb{W}(\mb{y}-\mb{Hx})\big\rangle\big\} \nonumber\\
		&\propto& p(x_i)\,\mr{exp} \big\{- \mb{h}_i^H\lr{\mb{W}}\mb{h}_i |x_i - z_i|^2\big\} \nonumber \\
		&\propto& p(x_i)\,\mc{CN}\big(z_i;x_i,{1}/{\big(\mb{h}_i^H\lr{\mb{W}}\mb{h}_i\big)} \big),
	\end{eqnarray}
	where $z_i$ is a linear estimate of $x_i$ that is now defined as
	\vspace{-0.15cm}
	\begin{eqnarray} \label{z-i-LMMSE-VB}
		z_i &=& \frac{\mb{h}^H_i\lr{\mb{W}}}{\mb{h}_i^H\lr{\mb{W}}\mb{h}_i}\bigg(\mb{y} - \sum_{j\neq i}^K \mb{h}_j\lr{x_j}\bigg) \nonumber \\
		&=& \lr{x_i} + \frac{\mb{h}^H_i\lr{\mb{W}}}{\mb{h}_i^H\lr{\mb{W}}\mb{h}_i}\big(\mb{y} - \mb{H}\lr{\mb{x}}\big).
	\end{eqnarray}
	The variational distribution $q_i(x_i)$ is realized by normalizing $p(x_i)\,\mc{CN}\big(z_i;x_i,1/(\mb{h}_i^H\lr{\mb{W}}\mb{h}_i)\big)$. The variational mean and variance are then computed as $\lr{x_i} = \ms{F}\big(z_i,1/(\mb{h}_i^H\lr{\mb{W}}\mb{h}_i)\big)$ and $\sigma_{x_i}^2 = \ms{G}\big(z_i,1/(\mb{h}_i^H\lr{\mb{W}}\mb{h}_i)\big)$, respectively. 
	
	\textit{2) Updating $\mb{W}$.} The variational distribution $q(\mb{W})$ is obtained by taking the expectation of the conditional in \eqref{factor-W} with respect to $q(\mb{x})$:
	\vspace{-0.15cm}
	\begin{eqnarray} \label{q-W}
		q(\mb{W}) &\propto& \mr{exp}\big\{\big\langle\ln p(\mb{y}|\mb{x},\mb{W};\mb{H}) + \ln p(\mb{W}) \big\rangle\big\}.
	\end{eqnarray}
	Observe that \eqref{q-W} can be expanded as in \eqref{W-distribution} at the top of the page by applying Corollary~\ref{corollary-1}. The variational distribution $q(\mb{W})$ is thus complex Wishart with $n+1$ degrees of freedom and mean $\lr{\mb{W}}$ given in \eqref{W-mean} at the top of the page. 
	
	By iteratively optimizing $\big\{q_i(x_i)\big\}$ and $q(\mb{W})$, we obtain the CAVI algorithm for estimating $\mb{x}$ and the precision matrix $\mb{W}$. We refer to this scheme as the \textbf{\textit{LMMSE-VB algorithm}} since $z_i$ resembles an LMMSE estimate of $x_i$ due to the cancellation of the inter-user interference and the whitening with the postulated noise covariance matrix $\mb{C}^{\mathrm{post}}$.
	
	\emph{\textbf{Remark 4:} If an improper prior $\mc{CW}(\mb{0},0)$ is used, the variational mean $\lr{\mb{W}}$ in \eqref{W-mean} cannot be computed due to the rank deficiency of $ \big(\mb{y}-\mb{H}\lr{\mb{x}}\big)\big(\mb{y}-\mb{H}\lr{\mb{x}}\big)^H+\mb{H}\mb{\Sigma}_{\mb{x}}\mb{H}^H$. In fact, it may not be possible to estimate the covariance matrix $\mb{C}^{\mathrm{post}} = \lr{\mb{W}}^{-1}$ with only one degree of freedom. To circumvent this issue, we propose to use the estimator
		\vspace{-0.15cm}
		\setcounter{equation}{40}
		\begin{eqnarray} \label{W-mean-2}
			\lr{\mb{W}} &\approx& \bigg( \frac{\big\|\mb{y}-\mb{H}\lr{\mb{x}}\big\|^2}{M}\mb{I}_M+\mb{H}\mb{\Sigma}_{\mb{x}}\mb{H}^H\bigg)^{-1}
		\end{eqnarray}
		for the precision matrix $\mb{W}$. Similar to the AMP-based algorithms and MF-VB, the term $\big(\big\|\mb{y}-\mb{H}\lr{\mb{x}}\big\|^2/M\big)\mb{I}_M$ reflects the empirical estimate of the true noise variance $N_0$ (and also guarantees the existence of the inverse), whereas the term $\mb{H}\mb{\Sigma}_{\mb{x}}\mb{H}^H$ reflects the empirical error covariance matrix induced by the MMSE denoiser $\ms{F}\big(z_i,1/\big(\mb{h}_i^H\lr{\mb{W}}\mb{h}_i\big)\big)$. Although the convergence of LMMSE-VB using $\lr{\mb{W}}$ in \eqref{W-mean-2} is not analytically proved, all the simulations presented in Section~\ref{sec:simu-Sect} indicate a robust and fast convergence as well as a remarkable performance.}
	
	\begin{algorithm}[t]
		\small
		\textbf{Input:} $\mb{y}$, $\mb{H}$, and prior distributions $\big\{p(x_i)\big\}$\;
		\textbf{Output:} $\mb{\hat{x}}$\; 
		Initialize $\hat{x}_i^1 = 0$ and $\sigma_{x_i,1}^{2}=\mr{Var}_{p(x_i)}[x_i],~\forall i$, and $\mb{r} = \mb{y} - \mb{H}\mb{\hat{x}}^1$\;
		\SetAlgoNoLine
		\For{$t=1,2,\ldots$}{
			Update $\bs{\Sigma}_{\mb{x}} = \mr{diag}(\sigma_{x_1,t}^{2},\ldots,\sigma_{x_K,t}^{2})$\;
			$\gamma_t \leftarrow M/\big(\|\mb{r}\|^2+\tr\{\mb{H}\mb{\Sigma}_{\mb{x}}\mb{H}^H\}\big)$ for MF-VB \emph{or} $\mb{W}_t \leftarrow \big((\|\mb{r}\|^2/M)\mb{I}_M+ \mb{H}\mb{\Sigma}_{\mb{x}}\mb{H}^H\big)^{-1}$ for LMMSE-VB\;
			\For{$i=1,2,\ldots,K$}{ 
				For MF-VB, compute:
				\begin{align*}
					z_i^t &\leftarrow \hat{x}_i^t + \mb{h}_i^H\mb{r}/\|\mb{h}_i\|^2, \\
					\hat{x}_i^{t+1} &\leftarrow \ms{F}\big(z_i^t, 1/(\gamma_t\|\mb{h}_i\|^2)\big), \\
					\sigma_{x_i,t+1}^{2} &\leftarrow \ms{G}\big(z_i^t, 1/(\gamma_t\|\mb{h}_i\|^2)\big),
				\end{align*}
				\emph{or}, for LMMSE-VB, compute:
				\begin{align*}
					z_i^t &\leftarrow \hat{x}_i^t + \mb{h}_i^H\mb{W}_t\mb{r}/(\mb{h}_i^H\mb{W}_t\mb{h}_i), \\
					\hat{x}_i^{t+1} &\leftarrow \ms{F}\big(z_i^t, 1/(\mb{h}_i^H\mb{W}_t\mb{h}_i)\big), \\
					\sigma_{x_i,t+1}^{2} &\leftarrow \ms{G}\big(z_i^t, 1/(\mb{h}_i^H\mb{W}_t\mb{h}_i)\big).
				\end{align*}\\
				Update residual: $\mb{r} \leftarrow \mb{r} + \mb{h}_i(\hat{x}_i^t - \hat{x}_i^{t+1})$
			}
		}
		MAP estimate: $\hat{x}_i \leftarrow \argmax_{a\in\mc{S}} p_a\mc{CN}\big(z_i^t;a, 1/(\gamma_t\|\mb{h}_i\|^2)\big)$ for MF-VB \emph{or}~$\argmax_{a\in\mc{S}} p_a\mc{CN}\big(z_i^t;a, 1/(\mb{h}_i^H\mb{W}_t\mb{h}_i)\big)$ for LMMSE-VB.
		\caption{\textit{\textbf{MF-VB/LMMSE-VB algorithm}} with postulated noise variance/covariance matrix}
		\label{algo-2}
	\end{algorithm}
	
	MF-VB and LMMSE-VB are summarized side by side in Algorithm~\ref{algo-2}. Here, we use $\hat{x}_i^t$ to replace the variational mean $\lr{x_i}$ at iteration~$t$ and each iteration consists of one round of updating $\{x_i\}$ and $\gamma$ (or $\mb{W}$). To reduce the complexity of both algorithms, we also include the residual term $\mb{r}$, which is initialized as $\mb{y} - \mb{H}\mb{\hat{x}}^1$. After re-estimating $x_i$, i.e., $\hat{x}_i^t$ into $\hat{x}_i^{t+1}$, the residual $\mb{r}$ is updated as in step~9 to reflect the update of $\hat{x}_i^{t+1}$. This step allows the matrix multiplication $\mb{H}\mb{\hat{x}}^t$ to be bypassed in the linear estimator to obtain $z_{i+1}^t$.
	
	\emph{\textbf{Remark 5:} MF-VB and LMMSE-VB are analogous to MF-SIC and LMMSE-SIC, respectively, except for a key difference. The reciprocal of the noise variance (or covariance matrix) in the VB algorithms is estimated only once per iteration. This implementation significantly reduces the complexity, especially for LMMSE-VB, compared with their SIC counterparts. In addition, the convergence of MF-VB to at least a local optimal solution can be analytically proved due to the coordinate ascent approach of the algorithm.}
	
	\emph{\textbf{Remark 6:} MF-VB and LMMSE-VB are similar to AMP and OAMP/VAMP implemented with the MF and the LMMSE filter, respectively, in the linear estimation step. However, the VB algorithms do not compute nor remove the divergence term as do their AMP-based counterparts. Another difference between the two frameworks lies in the updating step. The VB algorithms use successive updates (Gauss-Seidel method), in which each $z_i$ is computed based on the latest $\mb{\hat{x}}$ followed by the update of $\hat{x}_i$. On the other hand, the AMP-based schemes allow parallel updates (Jacobi method) of $\{z_i\}$ based on $\mb{\hat{x}}$ from the previous iteration followed by the update of $\{\hat{x}_i\}$. Interestingly, by using the residual update in step~9 of Algorithm~\ref{algo-2}, the complexity of each iteration of the VB methods becomes comparable to that of the AMP-based algorithms, as described next.}
	
	\vspace{-0.25cm}
	\subsection{Computational Complexity Analysis}
	
	This section presents a comparative analysis of the computational complexities of the considered detection algorithms assuming $M\geq K$ (as in an uplink MIMO system). These are summarized in Table \ref{table_complexities} and elaborated in the following. The complexities of the iterative algorithms (i.e., all but the LMMSE detector) are intended per iteration.
	
	The LMMSE detector has complexity $\mathcal{O}\big(MK^2 + |\mc{S}|K\big)$.
	AMP has complexity $\mc{O}\big(MK\big)$ in the linear estimation step and $\mc{O}\big(|\mc{S}|K\big)$ in the nonlinear denoiser for all the $K$ users. Due to the matrix inversion in the LMMSE filter, the complexity of OAMP/VAMP increases to $\mc{O}\big(MK^2+|\mc{S}|K\big)$. MF-SIC requires the computation of the residual interference-plus-noise covariance matrix $\mb{C}_i$ for each user~$i$, which has complexity $\mc{O}\big(M^2\big)$ with proper implementation. Thus, the total complexity of MF-SIC is $\mc{O}\big(M^2K+|\mc{S}|K\big)$. Due to the matrix inversion of $\mb{C}_i$ for each user~$i$ in the LMMSE filter, the complexity of LMMSE-SIC increases to $\mc{O}\big(M^3K+|\mc{S}|K\big)$. In conv-VB, the computation of the variational distribution $q_i(x_i)$ for each user~$i$ using $g_i(x_i)$ in \eqref{g-i} or via the expression of $z_i$ in \eqref{z-i} has complexity $\mc{O}\big(M^2+|\mc{S}|\big)$. However, by defining and properly updating the residual term $\mb{r} = \mb{y}-\mb{H}\mb{\hat{x}}$, the complexity can be reduced to $\mc{O}\big(M +|\mc{S}|\big)$ per user, resulting in the total complexity $\mc{O}\big(MK +|\mc{S}|K\big)$. Focusing on Algorithm~\ref{algo-2}, MF-VB has the same complexity as conv-VB with the use and update of $\mb{r}$ in step~9. Note that the update of $\gamma_t$ in step~6 requires the computation of $\tr\{\mb{H}\bs{\Sigma}_{\mb{x}}\mb{H}^H\} = \sum_{i=1}^K \|\mb{h}_i\|^2\sigma^2_{x_{i,t}}$, which itself has complexity $\mc{O}\big(MK\big)$. In LMMSE-VB, the computation of $\mb{W}_t$ in step~6 has complexity $\mc{O}\big(M^3\big)$, whereas the computation in step~9 has complexity $\mc{O}\big(M^2 + |\mc{S}|\big)$ per user. Hence, the total complexity of LMMSE-VB is $\mc{O}\big(M^3 + |\mc{S}|K\big)$.
	
	\begin{table}[t!]
		\centering
		\renewcommand{\arraystretch}{1.5}
		\centering
		\caption{Computational complexities.\label{table_complexities}}
		{\small 
			\begin{tabular}{|l|l|}
				\hline
				\textbf{Method} & \textbf{Complexity} \\
				\hline
				LMMSE & $\mathcal{O}\big(MK^2 + |\mc{S}|K\big)$ \\
				\hline
				AMP & $\mathcal{O}\big(MK + |\mc{S}|K \big)$ \\
				\hline
				OAMP/VAMP & $\mathcal{O}\big(MK^2 + |\mc{S}|K \big)$ \\
				\hline
				MF-SIC & $\mathcal{O}\big(M^2K + |\mc{S}|K \big)$ \\
				\hline
				LMMSE-SIC & $\mathcal{O}\big(M^3K + |\mc{S}|K \big)$ \\
				\hline
				conv-VB/MF-VB & $\mathcal{O}\big(MK + |\mc{S}|K\big)$ \\
				\hline
				LMMSE-VB &  $\mathcal{O}\big(M^3 + |\mc{S}|K\big)$ \\
				\hline
		\end{tabular}}
	\end{table}
	\vspace{-0.2cm}
	\section{VB Inference for MIMO Detection with Imperfect CSIR} \label{sec:VB_imperf}
	
	In this section, we develop a new VB method for MIMO detection in the presence of imperfect CSIR. We assume that there is a mismatch between the estimated channel, denoted by $\mb{\hat{H}}$, and the true channel $\mb{H}$. 
	
	\vspace{-0.2cm}
	\subsection{Conventional MIMO Detection with Imperfect CSIR}
	
	We first examine a conventional approach in which the BS estimates the uplink channel during the pilot transmission phase and uses the estimated channel for data detection. Let $\mb{x}_{\mr{p},i} \in \mathbb{C}^{T_{\mr{p}}}$ be the pilot transmitted by user~$i$. The received signal during the pilot transmission phase over $T_{\mr{p}}$ time slots can be modeled as 
	\vspace{-0.15cm}
	\begin{eqnarray}
		\mb{Y}_{\mr{p}} = \mb{H}\mb{X}_{\mr{p}} + \mb{N}_{\mr{p}},
	\end{eqnarray}
	where $\mb{X}_{\mr{p}} = [\mb{x}_{\mr{p},1},\ldots, \mb{x}_{\mr{p},K}]^T \in \mathbb{C}^{K\times T_{\mr{p}}}$ is the pilot matrix and $\mb{N}_{\mr{p}}$ is the AWGN with $\mc{CN}(0,N_0)$ elements. Here, we assume that the pilots of the $K$ users are orthogonal to each other, i.e., $\mb{X}_{\mr{p}}\mb{X}_{\mr{p}}^H = P_{\mr{p}} T_{\mr{p}}\mb{I}_K$, where $P_{\mr{p}}$ is the transmit power during the pilot transmission phase. We first correlate the received signal with the pilot of user~$i$ to obtain
	\vspace{-0.1cm}
	\begin{eqnarray}
		\mb{y}_{\mr{p},i} & =& \frac{1}{\sqrt{P_{\mr{p}}T_{\mr{p}}}} \mb{Y}_{\mr{p}} \mb{x}_{\mr{p},i}^*
		= \sqrt{P_{\mr{p}}T_{\mr{p}}} \mb{h}_i + \mb{{n}}_{\mr{p},i},
	\end{eqnarray} 
	where $\mb{{n}}_{\mr{p},i} = \big( 1/ \sqrt{P_{\mr{p}}T_{\mr{p}}}\big) \mb{N}_{\mr{p}}\mb{x}_{\mr{p},i}^* \sim \mc{CN}(0,N_0\mb{I}_M)$. 
	\begin{figure*}
		\setcounter{equation}{54}
		\begin{eqnarray} \label{x-dist}
			q_{i}(x_i) &\propto&p(x_i) \mr{exp}\bigg\{-\lr{\gamma} \big( \big\|\lr{\mb{h}_i}\big\|^2 + \tr\{\mb{\Sigma}_{i}\}\big)|x_i|^2 + 2\,\Re\bigg\{\lr{\mb{h}_i^H}\bigg(\mb{y} - \sum_{j\neq i}^K \lr{\mb{h}_j}\lr{x_j}\bigg)x_i^*\bigg\} \bigg\} \nonumber \\
			&\propto& p(x_i)\mr{exp}\Big\{-\lr{\gamma}\big(\big\| \lr{\mb{h}_i}\big\|^2 |x_i-z_i|^2 +\tr\{\mb{\Sigma}_{i}\}|x_i|^2\big)\Big\} \nonumber \\
			&\propto& p(x_i)\mc{CN}\Big(x_i;z_i,{1}/{\big(\lr{\gamma}\big\| \lr{\mb{h}_i}\big\|^2\big)}\Big)\mc{CN}\Big(x_i;0,{1}/{\big(\lr{\gamma}\tr\{\mb{\Sigma}_{i}\}\big)}\Big). 
		\end{eqnarray}
		\hrulefill
		\setcounter{equation}{43}
	\end{figure*} 
	Then, the optimal MMSE estimate $\mb{\hat{h}}_{i}$ can be obtained as
	\vspace{-0.1cm}
	\begin{eqnarray}\label{MMSE-channel-user}
		\mb{\hat{h}}_{i} &=& \mathbb{E}[\mb{h}_i\mb{y}_{\mr{p},i}^H] \big(\mathbb{E}[\mb{y}_{\mr{p},i}\mb{y}_{\mr{p},i}^H]\big)^{-1} \mb{y}_{\mr{p},i} \nonumber \\
		&=& (P_{\mr{p}}T_{\mr{p}} \mb{I}_M + N_0\mb{R}_i^{-1})^{-1}\mb{Y}_{\mr{p}}\mb{x}_{\mr{p},i}^*.
	\end{eqnarray}
	
	The estimation errors $\mb{e}_i = \mb{h}_i - \mb{\hat{h}}_i, \forall i$ are independent and each is distributed as $\mb{e}_i\sim \mc{CN}(\mb{0},\mb{K}_i)$, with
	\vspace{-0.1cm}
	\begin{eqnarray}
		\mb{K}_i = (P_{\mr{p}}T_{\mr{p}}N_0^{-1} \mb{I}_M +\mb{R}_i^{-1})^{-1}.
	\end{eqnarray} 
	The channel estimation mismatch can thus be modeled as
	\vspace{-0.1cm}
	\begin{eqnarray} 
		\mb{H} = \mb{\hat{H}} + \mb{E},
	\end{eqnarray}
	where $\mb{E}=[\mb{e}_1,\ldots,\mb{e}_K]$ is independent of the estimated channel $\mb{\hat{H}}$. Hence, the system model \eqref{system-model} can be rewritten as
	\vspace{-0.1cm}
	\begin{eqnarray}
		\mb{y} = \mb{\hat{H}}\mb{x} + \mb{E}\mb{x} + \mb{n}. 
	\end{eqnarray}
	Conditioned on $\mb{x}$, the effective noise $\mb{\tilde{n}} = \mb{E}\mb{x} + \mb{n}$ is Gaussian with zero mean and covariance matrix
	\vspace{-0.1cm}
	\begin{eqnarray}
		\mb{C}_{\mb{\tilde{n}}|\mb{x}} = \sum_{i=1}^K |x_i|^2\mb{K}_i + N_0\mb{I}_M.
	\end{eqnarray}
	Treating $p(\mb{y}|\mb{x};\mb{\hat{H}}) = \mc{CN}(\mb{y};\mb{\hat{H}}\mb{x},\mb{C}_{\mb{\tilde{n}}|\mb{x}})$ as the likelihood function, one can apply the MAP (or ML) detector as mentioned in Section~\ref{sec:conventional-detector} to obtain an optimal estimate of $\mb{x}$.
	
	Alternatively, an LMMSE detector can be used to estimate $\mb{x}$ with reduced complexity compared with the MAP detector. The LMMSE detector in \eqref{LMMSE} can be readily applied with a small adjustment by replacing the noise covariance matrix $N_0\mb{I}_M$ with the approximate covariance matrix of the effective noise $\mb{C}_{\mb{\tilde{n}}} = \mathbb{E}_{\mb{x}}\big[\mb{C}_{\mb{\tilde{n}}|\mb{x}}\big] = \sum_{i=1}^K \mathbb{E}\big[|x_i|^2\big]\mb{K}_i + N_0\mb{I}_M$. Note that conventional MIMO detection methods simply treat the channel estimation error as noise. Hence, we develop a novel VB scheme to jointly estimate the channel, the symbol vector, and the postulated noise variance.
	
	\vspace{-0.2cm}
	\subsection{Proposed MF-VB-M for MIMO Detection with Imperfect CSIR}
	
	Treating the channel $\mb{H}$ and the precision $\gamma=1/N_0^{\rm{post}}$ as random variables, the joint distribution $p(\mb{y}, \mb{x},\mb{H},\gamma; \mb{\hat{H}},\mb{K})$ can be factored as
	\vspace{-0.1cm}
	\begin{equation} \label{E-conditional}
		p(\mb{y}, \mb{x},\mb{H},\gamma; \mb{\hat{H}},\mb{K}) = p(\mb{y}| \mb{x},\mb{H},\gamma)p(\mb{H};\mb{\hat{H}},\mb{K})p(\mb{x})p(\gamma),
	\end{equation}
	where $p(\mb{y}| \mb{x},\mb{H},\gamma) = \mc{CN}(\mb{y};\mb{H}\mb{x},\gamma^{-1}\mb{I}_M)$ and $p(\mb{H};\mb{\hat{H}}, \mb{K}) \linebreak = \prod_{i=1}^K\mc{CN}(\mb{h}_i;\mb{\hat{h}}_i,\mb{K}_i)$. Given the observation $\mb{y}$ and the estimated channel $\mb{\hat{H}}$, we aim at obtaining the mean-field variational distribution $q(\mb{x},\mb{H}, \gamma)$ such that
	\vspace{-0.1cm}
	\begin{eqnarray}
		p(\mb{x},\mb{H},\gamma|\mb{y};\mb{\hat{H}},\mb{K}) &\approx& q(\mb{x},\mb{H},\gamma) \nonumber\\
		&=& \prod_{i=1}^K q_{i}(x_i) \prod_{i=1}^K q_{i}(\mb{h}_i)q(\gamma).
	\end{eqnarray}
	The optimization of $q(\mb{x},\mb{H},\gamma)$ is executed by iteratively updating $\{\mb{h}_i\}$, $\{x_i\}$, and $\gamma$ as follows.
	
	\textit{1) Updating $\mb{h}_i$.} The variational distribution $q_i(\mb{h}_i)$ is obtained by expanding the conditional in \eqref{E-conditional} and taking the expectation with respect to all latent variables except $\mb{h}_i$ using the variational distribution $q(\mb{x})\prod_{j\neq i}^K q_j(\mb{h}_j)q(\gamma)$:
	\vspace{-0.15cm}
	\begin{eqnarray} 
		\hspace{-7mm} && \hspace{-4mm} q_{i}(\mb{h}_i) \nonumber \\
		\hspace{-7mm} &\propto & \mr{exp}\big\{\big\langle\ln p(\mb{y}|\mb{x},\mb{H},\mb{\gamma}) + \ln p(\mb{h}_i;\mb{\hat{h}}_i,{\mb{K}}_{i}) \big\rangle\big\} \nonumber \\
		\hspace{-7mm} &\propto & \mr{exp}\big\{- \blr{\gamma \|\mb{y}-\mb{H}\mb{x}\|^2} - (\mb{h}_i-\mb{\hat{h}}_i)^H \mb{K}_i^{-1}(\mb{h}_i-\mb{\hat{h}}_i) \big\} \nonumber
		\\
		\hspace{-7mm} &\propto & \mr{exp}\bigg\{-\mb{h}_i^H\big[\lr{\gamma}\big\langle|x_i|^2\big\rangle\mb{I}_M+ \mb{K}_i^{-1}\big]\mb{h}_i \nonumber \\
		\hspace{-7mm} && +2\,\Re\bigg\{\mb{h}_i^H\bigg[\lr{\gamma}\bigg(\mb{y}\!-\! \sum_{j\neq i}^K \lr{\mb{h}_j}\lr{x_j}\!\bigg)\lr{x_i^*} \!+\! \mb{K}_i^{-1}\mb{\hat{h}}_i\bigg]\bigg\}\bigg\}.
	\end{eqnarray}
	The variational distribution $q_{i}(\mb{h}_i)$ is thus Gaussian with mean and covariance matrix
	\vspace{-0.1cm}
	\begin{align}
		\lr{\mb{h}_i} &= \mb{\Sigma}_{i}\bigg(\lr{\gamma}\bigg(\mb{y}- \sum_{j\neq i}^K \lr{\mb{h}_j}\lr{x_j}\bigg)\lr{x_i^*} + \mb{K}_i^{-1} \mb{\hat{h}}_i\bigg), \label{mean-e} \\
		\mb{\Sigma}_{i} &= \big[\lr{\gamma}\big\langle|x_i|^2\big\rangle\mb{I}_M + \mb{K}_i^{-1}\big]^{-1}, \label{Sigma-e}
	\end{align}
	respectively.
	
	\textit{2) Updating $x_i$.} The variational distribution $q_{i}(x_i)$ is obtained by expanding the conditional in \eqref{E-conditional} and taking the expectation with respect to all latent variables except $x_i$ using the variational distribution $\prod_{j\neq i}^K q_j(x_j)q(\mb{H})q(\gamma)$:
	\vspace{-0.1cm}
	\begin{eqnarray} \label{q_i}
		q_{i}(x_i)&\propto& \mr{exp}\big\{\big\langle\ln p(\mb{y}| \mb{x},\mb{H},\gamma)+\ln p(\mb{x})\big\rangle\big\} \nonumber \\
		&\propto& p(x_i)\mr{exp}\big\{-\big\langle\gamma \|\mb{y}-\mb{H}\mb{x}\|^2\big\rangle\big\}.
	\end{eqnarray}
	Note that \eqref{q_i} can be expanded as in \eqref{x-dist} at the top of the next page, where $z_i$ as a linear estimate of $x_i$ is defined as
	\setcounter{equation}{55}
	\vspace{-0.1cm}
	\begin{eqnarray} \label{z-imperfect}
		z_i &=&\frac{\lr{\mb{h}_i^H}}{\big\|\lr{\mb{h}_i}\big\|^2}\Bigg(\mb{y} - \sum_{j\neq i}^K \lr{\mb{h}_j}\lr{x_j}\Bigg) \nonumber \\
		&=& \lr{x_i} + \frac{\lr{\mb{h}_i^H}}{\big\|\lr{\mb{h}_i}\big\|^2}\big(\mb{y}-\lr{\mb{H}}\lr{\mb{x}}\big),
	\end{eqnarray}
	with $\big\langle\|\mb{h}_i\|^2\big\rangle = \big\|\lr{\mb{h}_i}\big\|^2 + \tr\{\mb{\Sigma}_{i}\}$. 
	Now, we define
	\vspace{-0.1cm}
	\begin{eqnarray} 
		\tilde{z}_i &=& \frac{z_i\big\|\lr{\mb{h}_i}\big\|^2}{\big\|\lr{\mb{h}_i}\big\|^2 +\tr\{\mb{\Sigma}_{i}\}}, \label{z-tilde}\\
		\tilde{\zeta}_i^2 &=& \frac{1}{\lr{\gamma}\big(\big\|\lr{\mb{h}_i}\big\|^2 +\tr\{\mb{\Sigma}_{i}\}\big)}, \label{sigma-tilde}
	\end{eqnarray}
	and obtain the variational distribution $q_i(x_i)$ as\footnote{To obtain \eqref{q-x-i}, we use the following property of the Gaussian distribution:
		\vspace{-0.15cm}
		\begin{eqnarray*}
			\mc{CN}(x;a,A)\mc{CN}(x;b,B) &=& \mc{CN}\left(x;\frac{a/A+b/B}{1/A+1/B},\frac{1}{1/A+1/B}\right)\nonumber \\
			&&\times\,\mc{CN}(0;a-b,A+B) \nonumber \\
			&\propto& \mc{CN}\left(x;\frac{a/A+b/B}{1/A+1/B},\frac{1}{1/A+1/B}\right).
	\end{eqnarray*}}
	\begin{eqnarray}\label{q-x-i}
		q_i(x_i) \propto p(x_i)\,\mc{CN}(x_i;\tilde{z}_i,\tilde{\zeta}_i^2),
	\end{eqnarray}
	which can be easily normalized. The variational mean $\lr{x_i}$ and variance $\sigma_{x_i}^2$ are then computed accordingly.

	\textit{3) Updating $\gamma$.} The variational distribution $q(\gamma)$ is obtained by taking the expectation of the conditional in \eqref{E-conditional} with respect to $q(\mb{x})q(\mb{H})$:
	\vspace{-0.15cm}
	\begin{eqnarray} 
		q(\gamma) &\propto& \mr{exp}\big\{\big\langle\ln p(\mb{y}|\mb{x},\mb{H},\mb{\gamma}) + \ln p(\gamma) \big\rangle\big\} \nonumber \\
		&\propto& \mr{exp}\big\{M\ln \gamma - \gamma \blr{\|\mb{y}-\mb{H}\mb{x}\|^2} \nonumber \\
		&& + (a_0-1)\ln\gamma - b_0\gamma\big\}.
	\end{eqnarray}
	The variational distribution $q(\gamma)$ is thus Gamma with mean
	\vspace{-0.15cm}
	\begin{eqnarray}\label{gamma-imperfect}
		\lr{\gamma} = \frac{a_0 + M}{b_0 + \blr{\|\mb{y}-\mb{H}\mb{x}\|^2}},
	\end{eqnarray}	
	to which we apply Theorem~\ref{theorem-1} to obtain
	\vspace{-0.15cm}
	\begin{eqnarray}\label{var-imperfect}
		\blr{\|\mb{y}-\mb{H}\mb{x}\|^2}&=& \big\|\mb{y}-\lr{\mb{H}}\lr{\mb{x}}\big\|^2 + \tr\big\{\lr{\mb{H}}\mb{\Sigma}_{\mb{x}}\lr{\mb{H}}^H\big\} \nonumber \\
		&& + \sum_{i=1}^K \big\langle|x_i|^2\big\rangle \tr\{\mb{\Sigma}_{i}\}. 
	\end{eqnarray}
	Similar to the AMP-based algorithms and MF-VB/LMMSE-VB, the term $\big\|\mb{y}-\lr{\mb{H}}\lr{\mb{x}}\big\|^2$ reflects the empirical estimate of the true noise variance $N_0$, whereas the term $\tr\big\{\lr{\mb{H}}\mb{\Sigma}_{\mb{x}}\lr{\mb{H}}^H\big\}$ reflects the empirical error covariance matrix induced by the MMSE denoiser $\ms{F}\big(\tilde{z}_i,\tilde{\zeta}_i^2\big)$. In addition, the term $\sum_{i=1}^K \big\langle|x_i|^2\big\rangle \tr\{\mb{\Sigma}_{i}\}$ reflects the empirical error covariance matrix induced by the channel estimator.
	
	By iteratively optimizing $\big\{q_i(\mb{h}_i)\big\}$, $\big\{q_i(x_i)\big\}$, and $q(\gamma)$, we obtain the CAVI algorithm for estimating $\mb{H}$, $\mb{x}$, and the precision $\gamma$. We refer to this scheme as the \textbf{\textit{MF-VB-M algorithm}} due to the use of the MF $\lr{\mb{h}_i}^H/\big\|\lr{\mb{h}_i}\big\|^2$ to obtain the linear estimate $z_i$ in \eqref{z-imperfect} with channel estimation mismatch. MF-VB-M is summarized in Algorithm~\ref{algo-3}. Here, we use $\hat{x}_i^t$ and $\mb{\check{h}}_i^t$ to replace the variational means $\lr{x_i}$ and $\lr{\mb{h}_i}$, respectively, at iteration~$t$ and each iteration consists of one round of updating $\{\mb{h}_i\}$, $\{x_i\}$, and $\gamma$.
	
	
	\begin{algorithm}[t]
		\small
		\textbf{Input:} $\mb{y}$, $\mb{\hat{H}}$, $\{\mb{K}_i\}$, and prior distributions $\big\{p(x_i)\big\}$\;
		\textbf{Output:} $\mb{\hat{x}}$ and $\mb{\check{H}}$\; 
		Initialize $\hat{x}_i^1 = 0$, $\sigma_{x_i,1}^{2}=\mr{Var}_{p(x_i)}[x_i]$, and $\bs{\Sigma}_i=\mb{K}_i,~\forall i$, and $\mb{\check{H}}^1 =\mb{\hat{H}}$\;
		\SetAlgoNoLine
		\For{$t=1,2,\ldots$}{
			\For{$i=1,2,\ldots,K$}{
				Compute $\mb{\check{h}}_i^t$ as in \eqref{mean-e} and $\bs{\Sigma}_i$ as in \eqref{Sigma-e}\;}
			\For{$i=1,2,\ldots,K$}{
				Compute $z_i^t$, $\tilde{z}_i^t$, and $\tilde{\zeta}_{i,t}^2$ as in \eqref{z-imperfect}, \eqref{z-tilde}, and \eqref{sigma-tilde}, respectively\;
				Compute and normalize $q_i(x_i)$ as in \eqref{q-x-i}\;
				Compute $\hat{x}_i^t$ and $\sigma_{x_i,t}^2$ with respect to $q_i(x_i)$\;}
			Compute $\gamma_t$ using \eqref{gamma-imperfect}--\eqref{var-imperfect}\;}
		MAP estimate: $\hat{x}\leftarrow \argmax_{a\in \mc{S}} q_i(a)$.
		\caption{\textit{\textbf{MF-VB-M algorithm}} with postulated noise variance and imperfect CSIR}
		\label{algo-3}
	\end{algorithm}
	
	\emph{\textbf{Remark 7:} Algorithm~\ref{algo-3} requires $K$ matrix inversions per iteration to compute the variational distribution $q(\mb{H})$. However, if the channel matrix $\mb{H}$ is i.i.d. Gaussian, i.e., $\mb{R}_i = (1/M)\mb{I}_M,~\forall i$, we obtain the statistics of the channel vector $\mb{h}_i$ during the pilot transmission phase as
		\begin{eqnarray*}
			\hat{\mb{h}}_i &=& \frac{1}{P_{\mr{p}} T_{\mr{p}} + MN_0}\mb{Y}_{\mr{p}} \mb{x}_{\mr{p},i}^*, \\
			\mb{K}_i &=& \frac{1}{P_{\mr{p}}T_{\mr{p}} N_0^{-1} + M}\mb{I}_M.
		\end{eqnarray*} 
		Algorithm~\ref{algo-3} can then be executed without any matrix inversion. More specifically, the variational distribution $q(\mb{h}_i)$ is Gaussian with mean and covariance matrix
		\begin{align}
			\lr{\mb{h}_i} & = \frac{\lr{\gamma}\big(\mb{y}- \sum_{j\neq i}^K \lr{\mb{h}_j}\lr{x_j}\big)\lr{x_i^*} + (P_{\mr{p}}T_{\mr{p}} N_0^{-1} + M) \mb{\hat{h}}_i}{\lr{\gamma} \big\langle|x_i|^2\big\rangle + P_{\mr{p}}T_{\mr{p}} N_0^{-1} + M}, \nonumber \\
			\bs{\Sigma}_i & = \frac{1}{\lr{\gamma} \big\langle |x_i|^2 \big\rangle + P_{\mr{p}}T_{\mr{p}} N_0^{-1} + M} \mb{I}_M, \nonumber
		\end{align}
		respectively.}
	
	\emph{\textbf{Remark 8:} LMMSE-VB can also be developed for the joint estimation of $\mb{H}$, $\mb{x}$, and the precision matrix $\mb{W} = \big(\mb{C}^{\mr{post}}\big)^{-1}$. This would call for a few minor adjustments to MF-VB-M to accommodate the estimation of $\mb{W}$ in place of $\gamma$ similar to those necessary to obtain LMMSE-VB from MF-VB.
		However, the computation of the variational distribution of $\mb{h}_i$, which is Gaussian with covariance matrix $\bs{\Sigma}_i = \big[\big\langle|x_i|^2\big\rangle\lr{\mb{W}} + \mb{K}_i^{-1}\big]^{-1}$, requires the inversion of an $(M\times M)$-dimensional matrix, even for i.i.d. channels. Hence, the resulting algorithm would be much more computationally intensive than LMMSE-VB. In addition, we observe through simulations that such an algorithm provides negligible performance gains with respect to MF-VB-M and we thus omit its derivations and discussion.}

	\section{Simulation Results} \label{sec:simu-Sect}
	
	This section presents numerical results comparing the SER performance of the AMP-based, SIC, and VB algorithms along with the LMMSE detector. The number of iterations is capped at $50$ for all the iterative algorithms. Unless otherwise stated, the covariance matrices $\{\mb{R}_i\}$ are normalized such that their diagonal elements are $1/M$, which implies $\mathbb{E}\big[\|\mb{h}_i\|^2\big] = 1,~\forall i$. The noise variance $N_0$ is set according to the operating signal-to-noise ratio (SNR), which is defined as
	\begin{equation}
		\mr{SNR}= \frac{\mathbb{E}\big[\|\mb{Hx}\|^2\big]}{\mathbb{E}\big[\|\mb{n}\|^2\big]}= \frac{\sum_{i=1}^K\tr\{\mb{R}_i\}}{MN_0} = \frac{K}{MN_0}.
	\end{equation}

	\subsection{Perfect CSIR with i.i.d. Gaussian Channels}
	
	We first examine the case where the channel matrix $\mb{H}$ consists of i.i.d. Gaussian coefficients (corresponding to i.i.d. Rayleigh fading) and is perfectly known at the BS.
	
\begin{figure}
		\includegraphics[width=80mm]{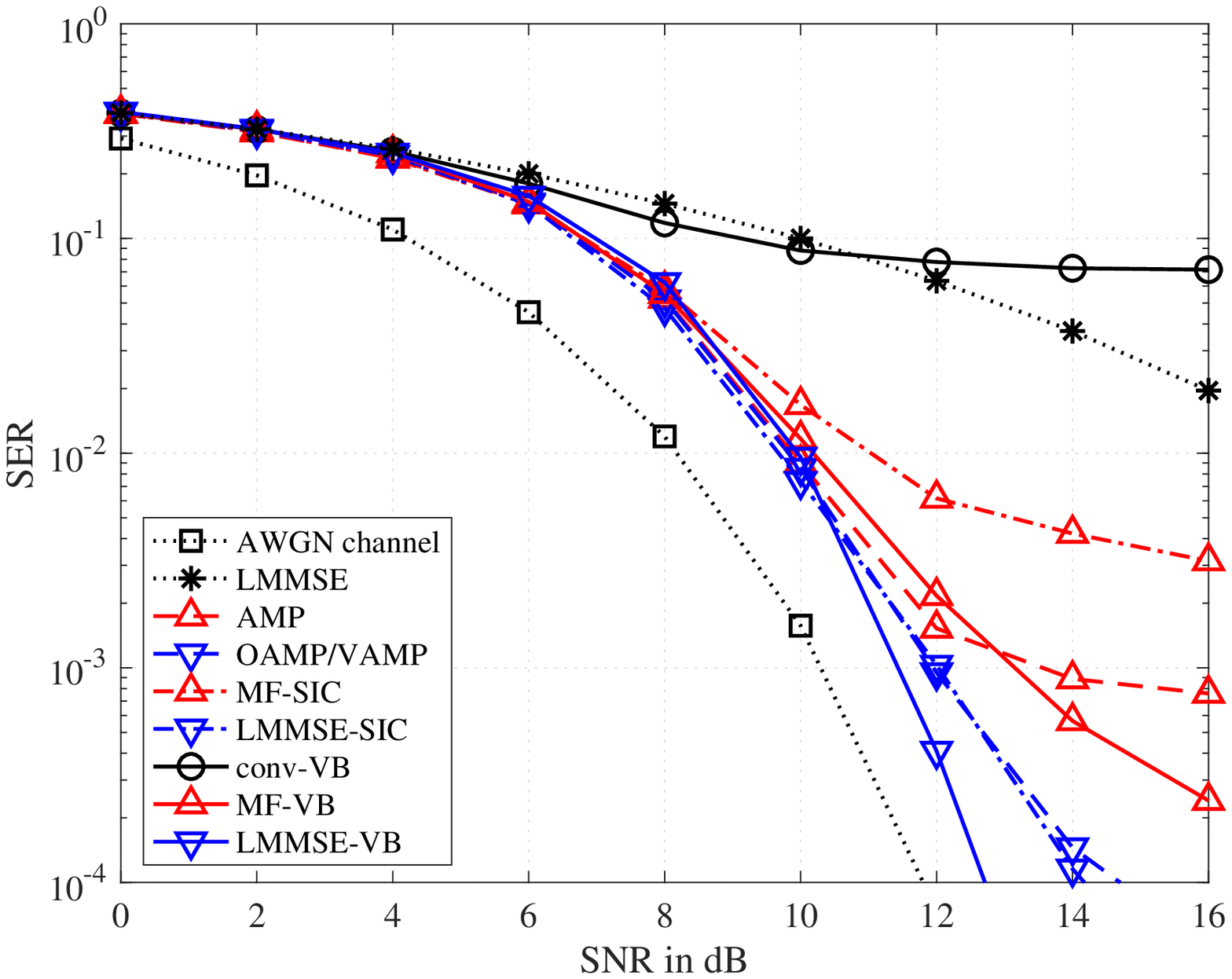}
		\caption{SER performance of the LMMSE detector, AMP-based algorithms (in \emph{dashed} lines), SIC algorithms (in \emph{dashed-dotted} lines), and VB algorithms (in \emph{solid} lines) assuming i.i.d. Rayleigh fading channels with $M=K=32$ and QPSK signaling. LMMSE-VB achieves the lowest SER at high SNR.}
		\label{fig-1}
		\includegraphics[width=80mm]{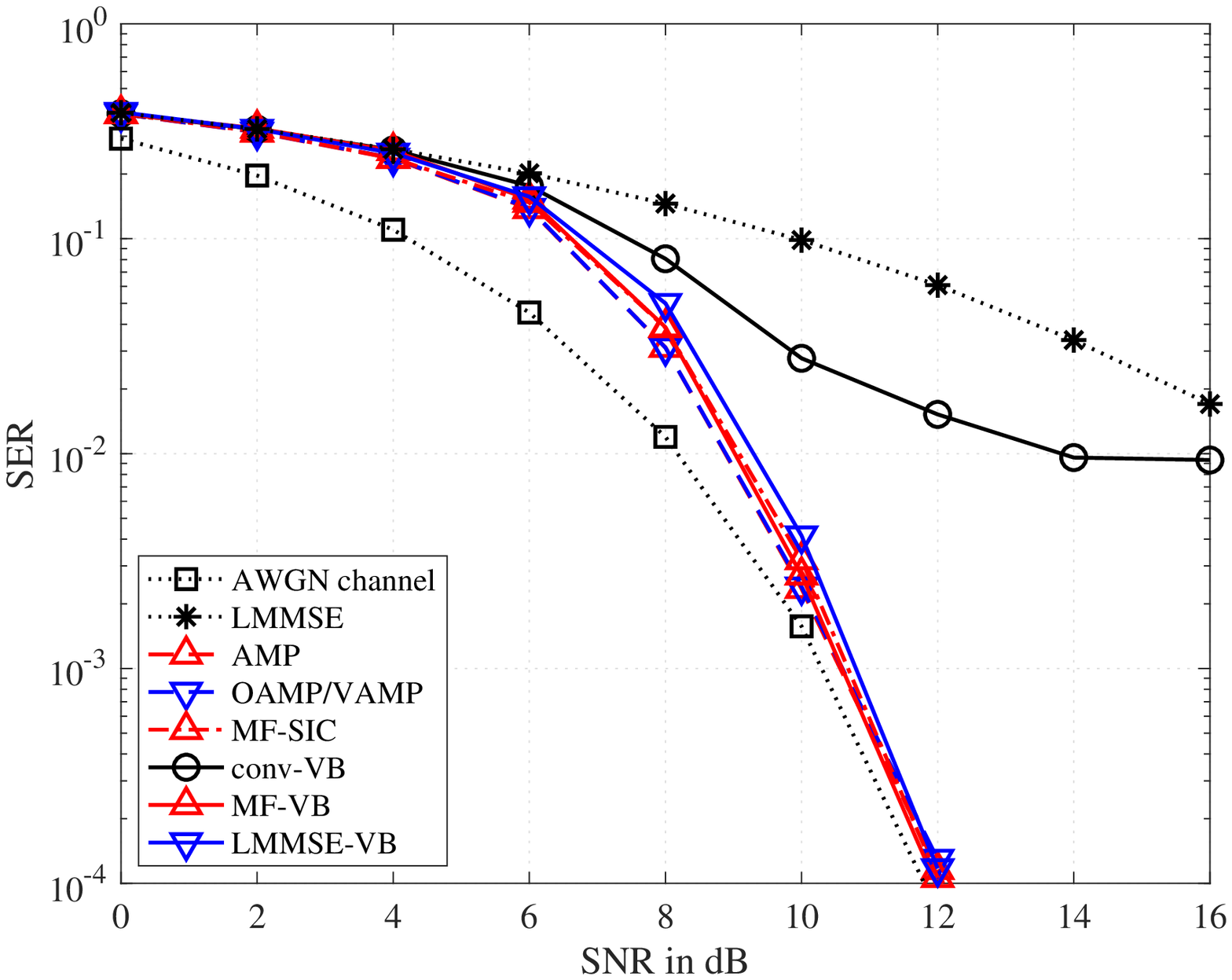}
		\caption{SER performance of the LMMSE detector, AMP-based algorithms (in \emph{dashed} lines), SIC algorithms (in \emph{dashed-dotted} lines), and VB algorithms (in \emph{solid} lines) assuming i.i.d. Rayleigh fading channels with $M=K=128$ and QPSK signaling. All the algorithms achieve comparable SER, except for the LMMSE detector and conv-VB.}
		\label{fig-2}
	\end{figure}

	Fig.~\ref{fig-1} illustrates the SER performance for a case with $M=K=32$ and QPSK signaling. At high SNR, MF-VB outperforms AMP and MF-SIC, whereas LMMSE-VB significantly outperforms OAMP/VAMP and LMMSE-SIC. In this relatively small MIMO system, the algorithms using the LMMSE filter in the linear estimation step significantly outperform their counterparts based on the MF. This gain comes at the cost of increased complexity, especially for LMMSE-SIC. It is observed that conv-VB performs very poorly, even worse than the LMMSE detector, at high SNR. The proposed MF-VB and LMMSE-VB have addressed this~limitation.
	
	Fig.~\ref{fig-2} depicts the SER performance for a case with $M=K=128$ and QPSK signaling. In this relatively large MIMO system, AMP, OAMP/VAMP, MF-SIC, MF-VB, and LMMSE-VB obtain similar SER results. Thus, in this case there is no benefit to use more computationally intensive schemes like OAMP/VAMP and LMMSE-VB. Due to the computational burden of LMMSE-SIC, we omit its simulation. However, LMMSE-SIC is expected to achieve a similar performance to the other algorithms (except conv-VB) since AMP and OAMP/VAMP are optimal in the large-system limit. In addition, as the residual inter-user interference becomes i.i.d. for large $K$ and i.i.d. channels, MF-SIC and LMMSE-SIC are thus equivalent. In Figs.~\ref{fig-1} and~\ref{fig-2}, the curve corresponding to AWGN channels is also plotted as a lower bound for i.i.d. Rayleigh fading channels with $\beta=1$ and $M\rightarrow \infty$.
	
\begin{figure}
		\includegraphics[width=80mm]{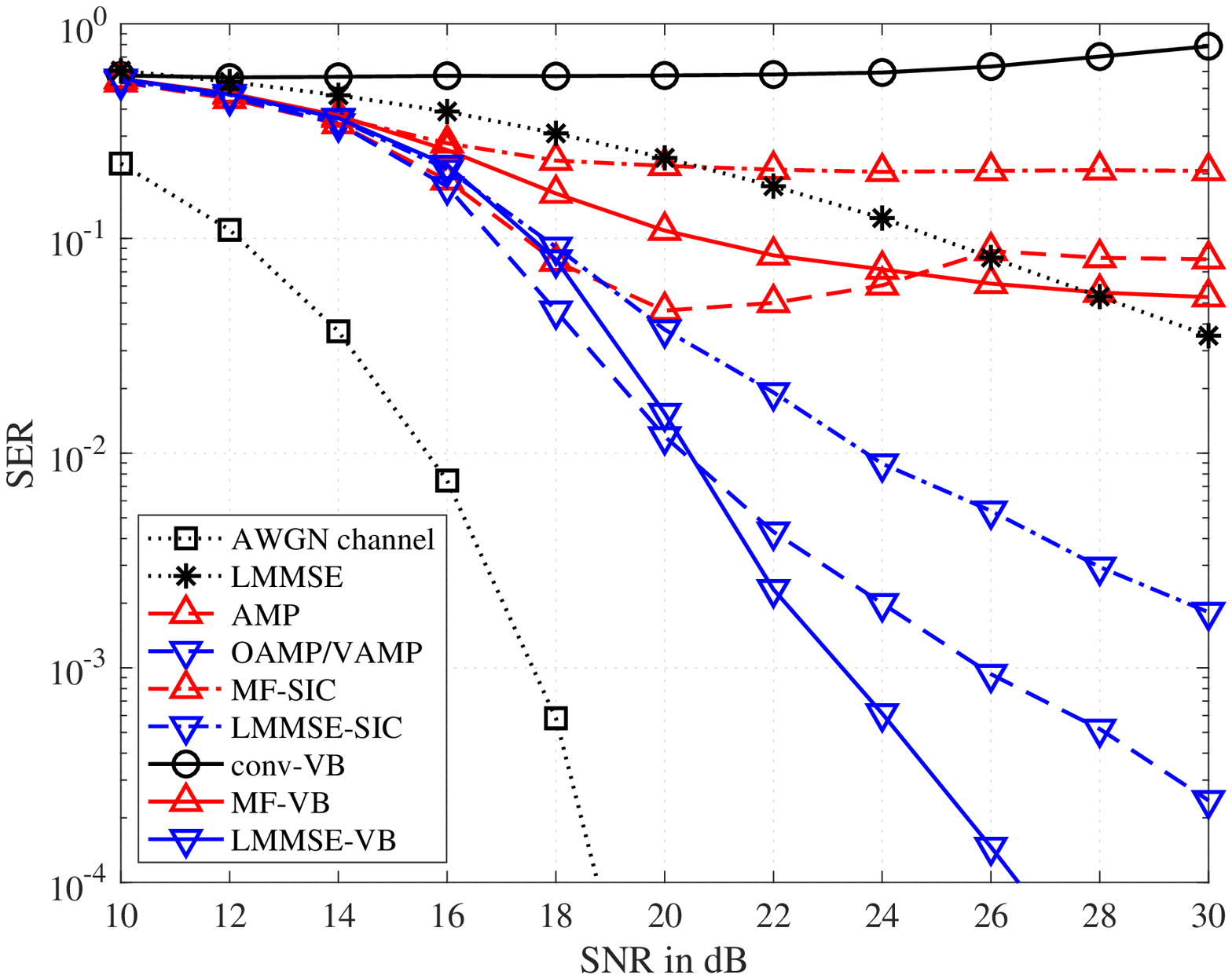}
		\caption{SER performance of the LMMSE detector, AMP-based algorithms (in \emph{dashed} lines), SIC algorithms (in \emph{dashed-dotted} lines), and VB algorithms (in \emph{solid} lines) assuming i.i.d. Rayleigh fading channels with $M=K=32$ and 16-QAM signaling. Only the algorithms using the LMMSE filter in the linear estimation step achieve acceptable SER, and LMMSE-VB shows the lowest SER at high SNR.}
		\label{fig-3}
		\includegraphics[width=80mm]{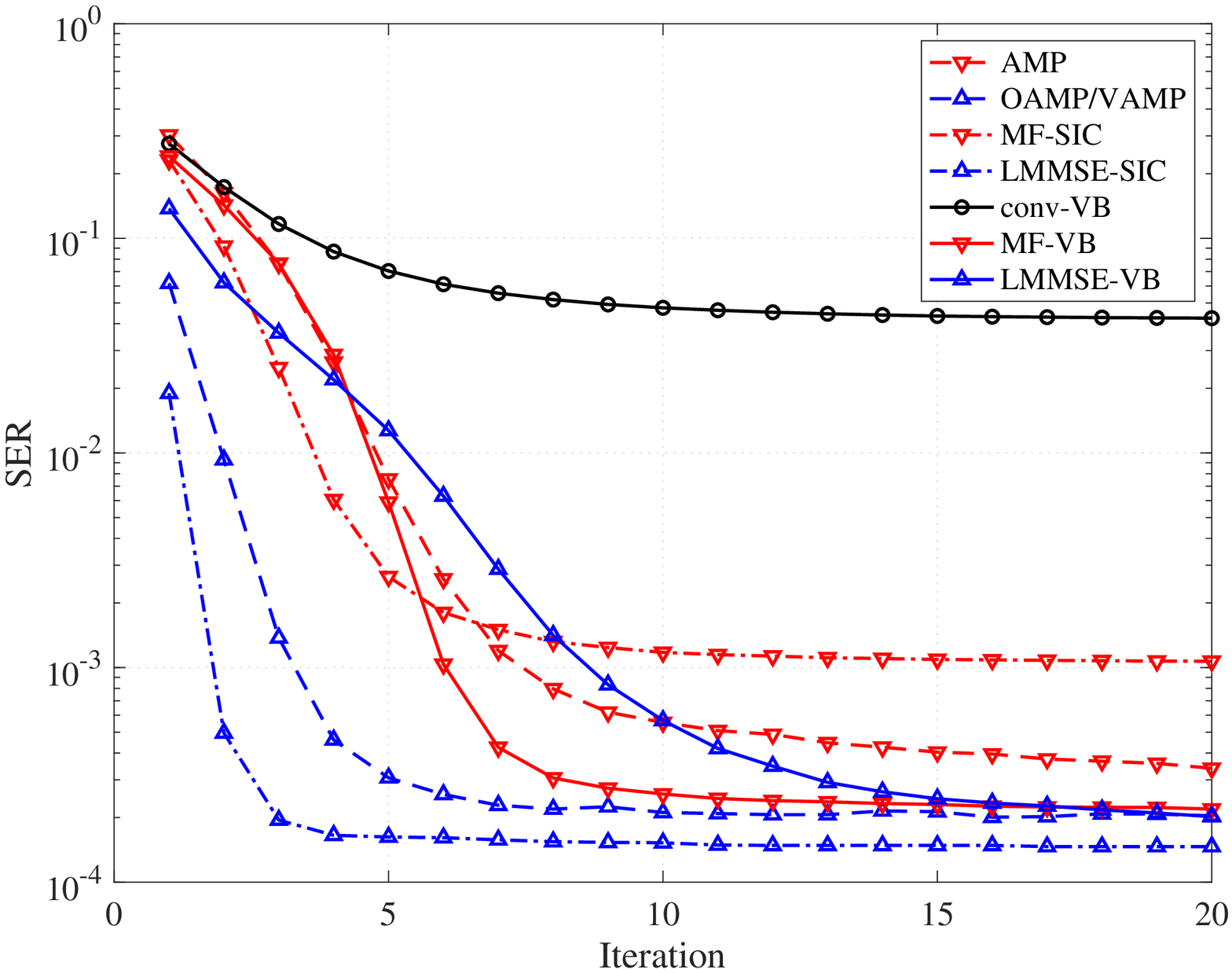}
		\caption{Convergence of the AMP-based algorithms (in \emph{dashed} lines), SIC algorithms (in \emph{dashed-dotted} lines), and VB algorithms (in \emph{solid} lines) assuming i.i.d. Rayleigh fading channels with $M=K=64$, QPSK signaling, and SNR of $12$~dB. All the algorithms exhibit very quick convergence (less than $20$ iterations).}
		\label{fig-4}
	\end{figure}

	Fig.~\ref{fig-3} plots the SER performance for a case with $M=K=32$ and 16-QAM signaling. Compared with the results in Fig.~\ref{fig-1}, it is worth noting that a higher-order modulation scheme requires more than a simple increase in SNR to approach the SER performance with AWGN channels. In this relatively small MIMO system, AMP, MF-SIC, and MF-VB saturate quite quickly and perform very poorly compared with OAMP/VAMP, LMMSE-SIC, and LMMSE-VB. Fig.~\ref{fig-3} also indicates the superior performance of the proposed LMMSE-VB over OAMP/VAMP and LMMSE-SIC, achieving gains of up to $5$~dB and $8$~dB, respectively, at high SNR.
	
	Fig.~\ref{fig-4} displays the convergence behavior of the iterative algorithms for a case with $M=K=64$, QPSK signaling, and an SNR of $12$~dB. The plots are obtained by averaging over $500$~channel realizations. The SIC algorithms converge faster than their AMP-based and VB counterparts. Specifically, LMMSE-SIC converges within $5$~iterations, whereas OAMP/VAMP requires $10$~iterations. However, the quick convergence of LMMSE-SIC comes at the cost of much higher complexity per iteration. 
	Interestingly, MF-VB converges faster and to a lower SER than AMP. Although its convergence is not analytically proved, LMMSE-VB converges fairly quickly in all the considered simulation scenarios.

	\subsection{Perfect CSIR with Correlated Channels}
	
	\begin{figure}
		\includegraphics[width=80mm]{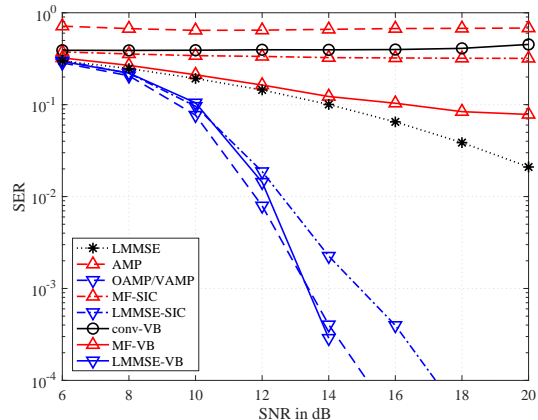}
		\caption{SER performance of the LMMSE detector, AMP-based algorithms (in \emph{dashed} lines), SIC algorithms (in \emph{dashed-dotted} lines), and VB algorithms (in \emph{solid} lines) assuming correlated Rayleigh fading channels based on the \emph{exponential model}, $M=K=64$, and QPSK signaling. Only the algorithms using the LMMSE filter in the linear estimation step achieve acceptable SER, and LMMSE-VB tends to achieve the lowest SER at high SNR.}
		\label{fig-5}
	\end{figure}

	\begin{figure}
		\vspace{-0.5cm}
		\includegraphics[width=80mm]{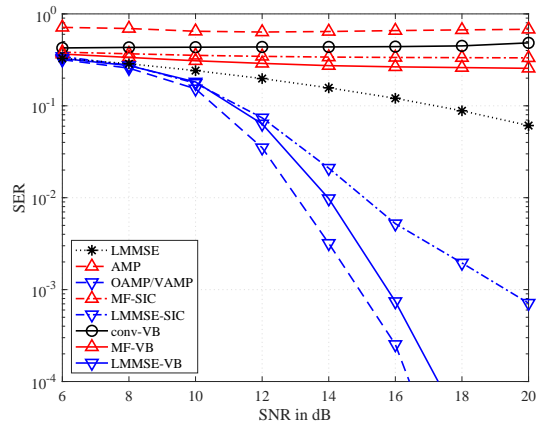}
		\caption{SER performance of the LMMSE detector, AMP-based algorithms (in \emph{dashed} lines), SIC algorithms (in \emph{dashed-dotted} lines), and VB algorithms (in \emph{solid} lines) assuming correlated Rayleigh fading channels based on the \emph{one-ring model}, $M=K=64$, and QPSK signaling. Only the algorithms using the LMMSE filter in the linear estimation step achieve acceptable SER, and OAMP/VAMP outperforms LMMSE-VB.}
		\label{fig-6}
	\end{figure}
	
	We now study the case where the channel matrix $\mb{H}$ consists of correlated Gaussian coefficients (corresponding to correlated Rayleigh fading) and is perfectly known at the BS.
	
	We first consider the exponential spatial correlation model~\cite{Loyka-CL-2001} for each column of $\mb{H}$, in which each covariance matrix $\mb{R}_i$ is set to
	\begin{eqnarray}
				[\mb{R}_i]_{k\ell} = \left\{\begin{array} {ll} 
					(1/M) \alpha^{k-\ell}, & \quad \textrm{ if } k\geq \ell \\
					(1/M) (\alpha^{\ell-k})^*, & \quad \textrm{ if } k< \ell, 
				\end{array} \right.
			\end{eqnarray}
			where $\alpha$ is the (complex) correlation coefficient between neighboring receive antennas. Fig.~\ref{fig-5} presents the SER performance with $M=K=64$, QPSK signaling, and using the exponential model with $\alpha=0.5+j0.5$. It is observed that only the algorithms using the LMMSE filter in the linear estimation step achieve acceptable SER at high SNR. AMP, MF-SIC, VB, and MF-VB are even worse than the LMMSE detector as they fail to account for the correlated MIMO channels in the linear estimation step. At very high SNR, LMMSE-VB tends to outperform OAMP/VAMP, and both achieve much lower SER than LMMSE-SIC.
			
			As a further example, we examine the SER performance using the one-ring spatial correlation model~\cite{Shiu-Foschini-TCOM-2020}. This is characterized by a ring of scatterers around the users and no significant local scattering around the BS. In this context, the multipath components arrive at the BS with a small angular spread and the covariance matrices $\{\mb{R}_i\}$ tend to have low rank as $M$ grows large~\cite{Adhikary-TIT-2013}. Fig.~\ref{fig-6} presents the SER performance for a case with $M=K=64$, QPSK signaling, and using the one-ring model with a $15^\circ$ angular spread. Similar to the results in Fig.~\ref{fig-5}, only OAMP/VAMP, LMMSE-SIC, and LMMSE-VB achieve acceptable SER at high SNR. However, OAMP/VAMP now outperforms LMMSE-VB by a small margin (i.e., $< 1$~dB).
			
			\begin{figure}
				\includegraphics[width=80mm]{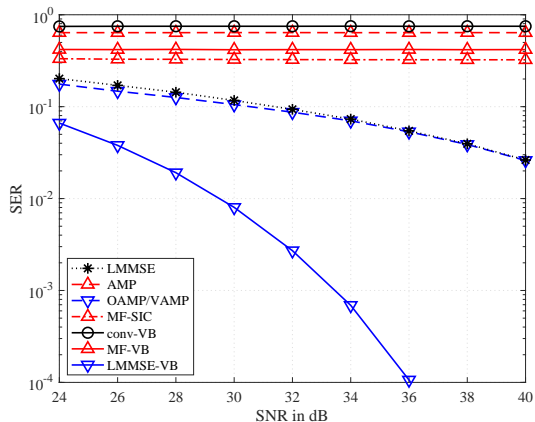}
				\caption{SER performance of the LMMSE detector, AMP-based algorithms (in \emph{dashed} lines), SIC algorithms (in \emph{dashed-dotted} lines), and VB algorithms (in \emph{solid} lines) assuming the \textit{QuaDRiGa channel simulator} with $M=128$, $K=64$, and QPSK signaling. LMMSE-VB outperforms all the other algorithms by a large margin.}
				\label{fig-7}
			\end{figure}
			
			We next consider a realistic channel model used to mimic urban cellular deployments. In particular, we assume 3D MIMO channels generated by the 3GPP QuaDRiGa channel simulator~\cite{Jaeckel2014Quadriga}. We consider a BS equipped with a rectangular planar array with $64$ dual-polarized antennas (i.e., $M=128$) installed at a height of $25$~m. The BS is assumed to cover a $120^\circ$ cell sector of radius $500$~m within which $K=64$ users are uniformly distributed. We generate $200$~channel realizations by creating $200$~independent realizations of the user locations. Since the pathloss can vary dramatically between different users and no power control is assumed ($\mathbb{E}[|x_i|^2]=1,~\forall i$), the operating SNRs can vary significantly from user to user. For each channel realization, we vary the noise variance $N_0$ at the BS accordingly to achieve an average operating SNR for all users, which is now defined as
			\begin{equation}
				\mr{SNR}= \frac{\mathbb{E}\big[\|\mb{Hx}\|^2\big]}{\mathbb{E}\big[\|\mb{n}\|^2\big]}= \frac{\tr\{\mb{H}\mb{H}^H\}}{MN_0}.
			\end{equation}
			Fig.~\ref{fig-7} illustrates the SER performance using the QuaDRiGa channel simulator described above, where LMMSE-SIC is omitted due to its prohibitive complexity with $M=128$. It is observed that OAMP/VAMP performs only slightly better than the LMMSE detector, and both are significantly worse than LMMSE-VB. The large gap between OAMP/VAMP and LMMSE-VB in this non-homogeneous SNR setting is due to the difference in their methods of decoupling the MIMO channel. OAMP/VAMP decouples the MIMO channel into $K$ parallel Gaussian channels with the same SNR, i.e., $z_i = x_i + \mc{CN}(0,\sigma_t^2)$. On the contrary, LMMSE-VB decouples it into $K$ parallel channels with possibly different SNRs, i.e., $z_i = x_i + \mc{CN}\big(0,1/(\mb{h}_i^H\mb{W}_t\mb{h}_i)\big)$, enabling the consideration of user-specific channel conditions.

			\subsection{Imperfect CSIR}
			
			\begin{figure}
				\includegraphics[width=80mm]{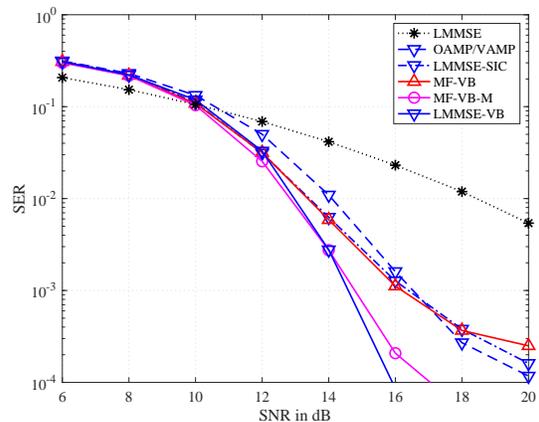}
				\caption{SER performance of the LMMSE detector, AMP-based algorithms (in \emph{dashed} lines), SIC algorithms (in \emph{dashed-dotted} lines), and VB algorithms (in \emph{solid} lines) assuming i.i.d. Rayleigh fading channels with imperfect CSIR, $M=K=32$, and QPSK signaling. MF-VB-M reduces the SER with respect to MF-VB by nearly one order of magnitude at high SNR.}
				\label{fig-8}
			\end{figure}
			
			Lastly, we examine the i.i.d. Rayleigh fading case with imperfect CSIR and compare the SER performance of the proposed MF-VB-M with that of MF-VB and of the other algorithms using the LMMSE filter in the linear estimation step. We consider $M=K=32$, QPSK signaling, pilot transmission time $T_{\mr{p}} = 32$, and pilot transmit power $P_{\mr{p}} = 1$~W. The estimated channel $\mb{\hat{H}}$ is obtained via the optimal MMSE channel estimator in \eqref{MMSE-channel-user}. In all the algorithms except MF-VB-M, the estimated channel $\mb{\hat{H}}$ is treated as the true channel $\mb{H}$. The resulting SER performance is illustrated in Fig.~\ref{fig-8}. Compared with the results in Fig.~\ref{fig-1}, it is observed that LMMSE-VB outperforms OAMP/VAMP and LMMSE-SIC by a wider margin in this case with channel estimation mismatch. Furthermore, the SER of MF-VB is close to that of OAMP/VAMP and LMMSE-SIC. The improved performance of MF-VB and LMMSE-VB relative to OAMP/VAMP and LMMSE-SIC is due to the fact that the postulated noise variance/covariance matrix implicitly takes into account the channel estimation error. The proposed MF-VB-M algorithm for MIMO detection with imperfect CSIR is much better than MF-VB and performs similarly to LMMSE-VB at high SNR. This performance gain only requires a few additional simple computation steps to derive the variational distribution of $\mb{h}_i$, as detailed in \emph{\textbf{Remark~7}}.

			\section{Conclusion} \label{sec:concl}
			
			This paper presented a study of massive MIMO detection from a variational Bayesian perspective. For the case of perfect CSIR, we developed the MF-VB and LMMSE-VB algorithms that use the noise variance and covariance matrix, respectively, postulated by the VB framework itself. These algorithms address the limitation in the conventional VB method with known noise variance and can approach and outperform their AMP-based and SIC counterparts in numerous channel settings. In addition, they involve closed-form and computationally efficient updates and exhibit very quick convergence. Finally, we proposed the MF-VB-M algorithm for the case of imperfect CSIR. Numerical results confirmed the superior performance of the developed VB algorithms over the AMP-based and SIC schemes, especially under correlated channels and imperfect CSIR. Future work may consider extensions to nonlinear, time-varying, and/or wideband MIMO channels.
			
			\appendices

			\section{Computation of $p(x_i|z_i,\sigma_t^2)$, $\ms{F}(z_i^t,\sigma_t^2)$, and $\ms{G}(z_i^t,\sigma_t^2)$} \label{sec:append-A}
			
			To compute the posterior mean and variance used in AMP and OAMP/VAMP, observe that the posterior distribution is given by
			\begin{eqnarray}
				p(x_i|z_i^t;\sigma_t^2) &=& \frac{1}{Z} p(z_i^t|x_i;\sigma_t^2)p(x_i) \nonumber \\
				&=& \frac{1}{Z} \mc{CN}(z_i^t; x_i, \sigma_t^2)p(x_i),
			\end{eqnarray}
			where $Z$ is a normalization factor. In the context of MIMO detection, the posterior distribution is discrete with probability mass function given by
			\begin{eqnarray}
				p(a|z_i^t;\sigma_t^2) = \frac{1}{Z} \mr{exp}\bigg(-\frac{|z_i^t-a|^2 }{\sigma_t^2}\bigg)p_a,
			\end{eqnarray}
			with $Z = {\sum_{b\in \mc{S}} \mr{exp}\big(-\frac{|z_i^t-b|^2 }{\sigma_t^2}\big)p_b}$. The corresponding posterior mean $\ms{F}(z^t_i,\sigma_t^2)$ and variance $\ms{G}(z^t_i,\sigma_t^2)$ can be computed accordingly. The final MAP estimate of $x_i$ can be obtained as
			\begin{eqnarray}\label{MAP-estimate}
				\hat{x}_i &=& \argmax_{x_i\in \mc{S}}\; p(x_i|z_i^t;\sigma_t^2) \nonumber \\
				&=& \argmax_{a\in \mc{S}}\left(\ln p_a - \frac{|z_i^t - a|^2}{\sigma_t^2} \right).
			\end{eqnarray}
			


\begin{thebibliography}{10}
	\providecommand{\url}[1]{#1}
	\csname url@samestyle\endcsname
	\providecommand{\newblock}{\relax}
	\providecommand{\bibinfo}[2]{#2}
	\providecommand{\BIBentrySTDinterwordspacing}{\spaceskip=0pt\relax}
	\providecommand{\BIBentryALTinterwordstretchfactor}{4}
	\providecommand{\BIBentryALTinterwordspacing}{\spaceskip=\fontdimen2\font plus
		\BIBentryALTinterwordstretchfactor\fontdimen3\font minus
		\fontdimen4\font\relax}
	\providecommand{\BIBforeignlanguage}[2]{{%
			\expandafter\ifx\csname l@#1\endcsname\relax
			\typeout{** WARNING: IEEEtran.bst: No hyphenation pattern has been}%
			\typeout{** loaded for the language `#1'. Using the pattern for}%
			\typeout{** the default language instead.}%
			\else
			\language=\csname l@#1\endcsname
			\fi
			#2}}
	\providecommand{\BIBdecl}{\relax}
	\BIBdecl
	
	\bibitem{LuLu-JSAC-2014}
	L.~Lu, G.~Y. Li, A.~L. Swindlehurst, A.~Ashikhmin, and R.~Zhang, ``An overview
	of massive {MIMO}: Benefits and challenges,'' \emph{IEEE J. Sel. Areas
		Commun.}, vol.~8, no.~5, pp. 742--758, Oct. 2014.
	
	\bibitem{Rusek-SPMag-2013}
	F.~Rusek, D.~Persson, B.~K. Lau, E.~G. Larsson, T.~L. Marzetta, O.~Edfors, and
	F.~Tufvesson, ``Scaling up {MIMO:} {O}pportunities and challenges with very
	large arrays,'' \emph{IEEE Signal Process. Mag.}, vol.~30, no.~1, pp. 40--60,
	Jan. 2013.
	
	\bibitem{Yang-Hanzo-CST}
	S.~Yang and L.~Hanzo, ``Fifty years of {MIMO} detection: The road to
	large-scale {MIMOs},'' \emph{IEEE Commun. Surveys and Tutorials}, vol.~17,
	no.~4, pp. 1941--1988, 2015.
	
	\bibitem{Wang-TCOM-1999}
	X.~Wang and H.~Poor, ``Iterative (turbo) soft interference cancellation and
	decoding for coded {CDMA},'' \emph{IEEE Trans. Commun.}, vol.~47, no.~7, pp.
	1046--1061, Jul. 1999.
	
	\bibitem{Alexander-TCOM-1999}
	P.~Alexander, M.~Reed, J.~Asenstorfer, and C.~Schlegel, ``Iterative multiuser
	interference reduction: {T}urbo {CDMA},'' \emph{IEEE Trans. Commun.},
	vol.~47, no.~7, pp. 1008--1014, Jul. 1999.
	
	\bibitem{Choi-Cioffi-WCNC-2000}
	W.-J. Choi, K.-W. Cheong, and J.~Cioffi, ``Iterative soft interference
	cancellation for multiple antenna systems,'' in \emph{Proc. IEEE Wireless
		Commun. and Netw. Conf. (WCNC)}, Sep. 2000.
	
	\bibitem{Shlezinger-DeepSIC-TWC-2021}
	N.~Shlezinger, R.~Fu, and Y.~C. Eldar, ``{DeepSIC}: Deep soft interference
	cancellation for multiuser {MIMO} detection,'' \emph{IEEE Trans. Wireless
		Commun.}, vol.~20, no.~2, pp. 1349--1362, Feb. 2021.
	
	\bibitem{Donoho}
	D.~L. Donoho, A.~Maleki, and A.~Montanari, ``Message-passing algorithms for
	compressed sensing,'' \emph{Proc. Nat. Academy Sci.}, vol. 106, no.~45, pp.
	18\,914--18\,919, Nov. 2009.
	
	\bibitem{Jeon-Studer}
	C.~Jeon, R.~Ghods, A.~Maleki, and C.~Studer, ``Optimality of large {MIMO}
	detection via approximate message passing,'' in \emph{Proc. IEEE Int. Symp.
		Inf. Theory (ISIT)}, Jun. 2015.
	
	\bibitem{Bayati-TIT-2011}
	M.~Bayati and A.~Montanari, ``The dynamics of message passing on dense graphs,
	with applications to compressed sensing,'' \emph{IEEE Trans. Inf. Theory},
	vol.~57, no.~2, pp. 764--785, Feb. 2011.
	
	\bibitem{Bayati-Annal-2014}
	M.~Bayati, M.~Lelarge, and A.~Montanari, ``{Universality in polytope phase
		transitions and message passing algorithms},'' \emph{Ann. Appl. Probability},
	vol.~25, no.~2, pp. 753--822, Apr. 2015.
	
	\bibitem{OAMP-2017}
	J.~{Ma} and L.~{Ping}, ``Orthogonal {AMP},'' \emph{{IEEE Access}}, vol.~5, pp.
	2020--2033, Jan. 2017.
	
	\bibitem{VAMP}
	S.~Rangan, P.~Schniter, and A.~K. Fletcher, ``Vector approximate message
	passing,'' \emph{IEEE Trans. Inf. Theory}, vol.~65, no.~10, pp. 6664--6684,
	May 2019.
	
	\bibitem{Thoota-TCOM-2021}
	S.~S. Thoota and C.~R. Murthy, ``Variational {B}ayes' joint channel estimation
	and soft symbol decoding for uplink massive {MIMO} systems with low
	resolution {ADCs},'' \emph{IEEE Trans. Commun.}, vol.~69, no.~5, pp.
	3467--3481, May 2021.
	
	\bibitem{Bishop-2006}
	C.~M. Bishop, \emph{Pattern recognition and machine learning}.\hskip 1em plus
	0.5em minus 0.4em\relax New York, NY, USA: Springer, 2006.
	
	\bibitem{Bjornson-TIT-2014}
	E.~{Bj{\"o}rnson}, J.~{Hoydis}, M.~{Kountouris}, and M.~{Debbah}, ``Massive
	{MIMO} systems with non-ideal hardware: Energy efficiency, estimation, and
	capacity limits,'' \emph{IEEE Trans. Inf. Theory}, vol.~60, no.~11, pp.
	7112--7139, Nov. 2014.
	
	\bibitem{Rangan-2011}
	S.~Rangan, ``Generalized approximate message passing for estimation with random
	linear mixing,'' in \emph{Proc. IEEE Int. Symp. Inf. Theory (ISIT)}, Jul.
	2011.
	
	\bibitem{Takeuchi-TIT-2021}
	K.~Takeuchi, ``Bayes-optimal convolutional {AMP},'' \emph{IEEE Trans. Inf.
		Theory}, vol.~67, no.~7, pp. 4405--4428, May 2021.
	
	\bibitem{Wainwright-2008}
	M.~J. Wainwright and M.~I. Jordan, ``Graphical models, exponential families,
	and variational inference,'' \emph{Found. and Trends\textregistered~Mach.
		Learn.}, vol.~1, no. 1--2, pp. 1--305, Jan. 2008.
	
	\bibitem{Krzakala-ISIT-2014}
	F.~Krzakala, A.~Manoel, E.~W. Tramel, and L.~Zdeborov{\'a}, ``Variational free
	energies for compressed sensing,'' in \emph{Proc. IEEE Int. Symp. Inf. Theory
		(ISIT)}, Aug. 2014.
	
	\bibitem{Feller-1968}
	W.~Feller, \emph{An Introduction to Probability Theorey and Its
		Applications}.\hskip 1em plus 0.5em minus 0.4em\relax New York, NY, USA:
	Wiley, 1968.
	
	\bibitem{Loyka-CL-2001}
	S.~Loyka, ``Channel capacity of {MIMO} architecture using the exponential
	correlation matrix,'' \emph{IEEE Commun. Letters}, vol.~5, no.~9, pp.
	369--371, Sep. 2001.
	
	\bibitem{Shiu-Foschini-TCOM-2020}
	D.-S. Shiu, G.~Foschini, M.~Gans, and J.~Kahn, ``Fading correlation and its
	effect on the capacity of multielement antenna systems,'' \emph{IEEE Trans.
		Commun.}, vol.~48, no.~3, pp. 502--513, Mar. 2000.
	
	\bibitem{Adhikary-TIT-2013}
	A.~Adhikary, J.~Nam, J.~Y. Ahn, and G.~Caire, ``Joint spatial division and
	multiplexing -- {T}he large-scale array regime,'' \emph{IEEE Trans. Inf.
		Theory}, vol.~59, no.~10, pp. 6441--6463, Oct. 2013.
	
	\bibitem{Jaeckel2014Quadriga}
	S.~Jaeckel, L.~Raschkowski, K.~B{\"o}rner, and L.~Thiele, ``Qua{DR}i{G}a: {A}
	3-{D} multi-cell channel model with time evolution for enabling virtual field
	trials,'' \emph{IEEE Trans. Antennas Propag}, vol.~62, no.~6, pp. 3242--3256,
	Jun. 2014.
	
\end{thebibliography}
		\end{document}